\documentclass[5p, times]{elsarticle}
\usepackage[utf8]{inputenc}

\usepackage{amssymb}
\usepackage{amsthm}
\usepackage{amsmath}
\usepackage{lineno}
\usepackage{xcolor}
\usepackage{lipsum}
\usepackage{enumitem}

\newcommand{\Eqref}[1]{Eq.~\eqref{#1}}
\newcommand{\qfact}{F_\mathrm{Q}}

\begin{document}

\begin{frontmatter}

\title{Design, Construction, and Test of the Gas Pixel Detectors for the IXPE Mission}
\author[1,2]{Baldini L.}
\author[3,2]{Barbanera M.}
\author[2]{Bellazzini R.}
\author[4,5]{Bonino R.}
\author[5]{Borotto F.}
\author[2]{Brez A.}
\author[5]{Caporale C.}
\author[2]{Cardelli C.}
\author[2]{Castellano S.}
\author[2]{Ceccanti M.}
\author[2]{Citraro S.}
\author[6]{Di Lalla N.}
\author[5]{Latronico L.}
\author[2]{Lucchesi L.}
\author[2]{Magazzù C.}
\author[2]{Magazzù G.}
\author[5]{Maldera S.}
\author[2]{Manfreda A.}
\author[5]{Marengo M.}
\author[2]{Marrocchesi A.}
\author[5]{Mereu P.}
\author[2]{Minuti M.}
\author[5]{Mosti F.}
\author[3,2]{Nasimi H.}
\author[2]{Nuti A.}
\author[5]{Oppedisano C.}
\author[2]{Orsini L.}
\author[2]{Pesce-Rollins M.}
\author[2]{Pinchera M.}
\author[2]{Profeti A.}
\author[2]{Sgrò C.\corref{corr}}
\author[2]{Spandre G.}
\author[5]{Tardiola M.}
\author[2]{Zanetti D.}
\author[7]{Amici F.}
\author[8]{Andersson H.}
\author[9]{Attinà P.}
\author[10]{Bachetti M.}
\author[11]{Baumgartner W.}
\author[7]{Brienza D.}
\author[12]{Carpentiero R.}
\author[12]{Castronuovo M.}
\author[13]{Cavalli L.}
\author[12]{Cavazzuti E.}
\author[14]{Centrone M.}
\author[7]{Costa E.}
\author[13]{D'Alba E.}
\author[12]{D'Amico F.}
\author[7]{Del Monte E.}
\author[7]{Di Cosimo S.}
\author[7]{Di Marco A.}
\author[7]{Di Persio G.}
\author[12]{Donnarumma I.}
\author[7]{Evangelista Y.}
\author[7]{Fabiani S.}
\author[7,15,16]{Ferrazzoli R.}
\author[17]{Kitaguchi T.}
\author[7]{La Monaca F.}
\author[7]{Lefevre C.}
\author[7]{Loffredo P.}
\author[13]{Lorenzi P.}
\author[13]{Mangraviti E.}
\author[18]{Matt G.}
\author[8]{Meilahti T.}
\author[7]{Morbidini A.}
\author[7]{Muleri F.}
\author[17]{Nakano T.}
\author[12]{Negri B.}
\author[8]{Nenonen S.}
\author[11]{O'Dell S. L.}
\author[14]{Perri M.}
\author[7]{Piazzolla R.}
\author[13]{Pieraccini S.}
\author[10]{Pilia M.}
\author[12]{Puccetti S.}
\author[11]{Ramsey B. D.}
\author[7,15,16]{Rankin J.}
\author[7,15,16]{Ratheesh A.}
\author[7]{Rubini A.}
\author[7]{Santoli F.}
\author[13]{Sarra P.}
\author[7]{Scalise E.}
\author[13]{Sciortino A.}
\author[7]{Soffitta P.}
\author[17]{Tamagawa T.}
\author[11]{Tennant A. F.}
\author[7]{Tobia A.}
\author[10]{Trois A.}
\author[17]{Uchiyama K.}
\author[13]{Vimercati M.}
\author[11]{Weisskopf M. C.}
\author[7]{Xie F.}
\author[13]{Zanetti F.}
\author[17]{Zhou Y.}
\cortext[corr]{Corresponding author}
\address[1]{Università di Pisa, Dipartimento di Fisica Enrico Fermi, Largo B. Pontecorvo 3, I-56127 Pisa, Italy}
\address[2]{Istituto Nazionale di Fisica Nucleare, Sezione di Pisa, Largo B. Pontecorvo 3, I-56127 Pisa, Italy}
\address[3]{Università di Pisa, Dipartimento di Ingegneria dell'Informazione, Via G. Caruso 16, I-56122 Pisa, Italy}
\address[4]{Università di Torino, Dipartimento di Fisica, Via P. Giuria 1, I-10125 Torino, Italy}
\address[5]{Istituto Nazionale di Fisica Nucleare, Sezione di Torino, Via P. Giuria, 1, I-10125 Torino, Italy}
\address[6]{W.W. Hansen Experimental Physics Laboratory, Kavli Institute for Particle Astrophysics and Cosmology, Department of Physics and SLAC National AcceleratorLaboratory, Stanford University, Stanford, CA 94305, USA}
\address[7]{Istituto di Astrofisica e Planetologia Spaziali di Roma, Via Fosso del Cavaliere 100, I-00133 Roma, Italy}
\address[8]{Oxford Instruments Technologies  Oy, Technopolis Innopoli 1, Tekniikantie 12, FI-02150 Espoo}
\address[9]{INAF/Osservatorio Astrofisico di Torino, Via Osservatorio 20, I-10025 Pino Torinese (TO), Italy}
\address[10]{INAF/Osservatorio Astronomico di Cagliari, Via della Scienza 5, I-09047 Selargius (CA), Italy}
\address[11]{NASA Marshall Space Flight Center, Huntsville, AL 35812, USA}
\address[12]{Agenzia Spaziale Italiana, Via del Politecnico snc, I-00133 Roma, Italy}
\address[13]{Orbitale Hochtechnologie Bremen, OHB Italia, Via Gallarate 150, I-20151 Milano, Italy}
\address[14]{INAF/Osservatorio Astronomico di Roma, Via Frascati 33, I-00040, Monte Porzio Catone (RM)}
\address[15]{Università di Roma Sapienza, Dipartimento di Fisica, Piazzale Aldo Moro 2, 00185 Roma, Italy}
\address[16]{Università di Roma Tor Vergata, Dipartimento di Fisica, Via della Ricerca Scientifica, 1, 00133 Roma, Italy}
\address[17]{RIKEN Nishina Center, 2-1 Hirosawa, Wako, Saitama 351-0198, Japan}
\address[18]{Università Roma Tre, Dipartimento di Matematica e Fisica, Via della Vasca Navale 84, I-00146, Italy}

\journal{Astroparticle Physics}
\date{Compiled on \today}
\fntext[cclicense]{© 2021. This manuscript version is made available under the CC-BY-NC-ND 4.0 license  
http://creativecommons.org/licenses/by-nc-nd/4.0/}

\begin{abstract}
Due to be launched in late 2021, the Imaging X-Ray Polarimetry Explorer (IXPE)
is a NASA Small Explorer mission designed to perform polarization measurements
in the 2--8~keV band, complemented with imaging, spectroscopy and timing
capabilities. At the heart of the focal plane is a set of three
polarization-sensitive Gas Pixel Detectors (GPD), each based on a custom ASIC acting as
a charge-collecting anode.

In this paper we shall review the design, manufacturing, and test of the IXPE
focal-plane detectors, with particular emphasis on the connection between the 
science drivers, the performance metrics and the operational aspects. We shall
present a thorough characterization of the GPDs in terms of effective noise,
trigger efficiency, dead time, uniformity of response, and spectral and
polarimetric performance. In addition, we shall discuss in detail a number of
instrumental effects that are relevant for high-level science analysis---particularly
as far as the response to unpolarized radiation and the stability in time are
concerned.
\end{abstract}


\end{frontmatter}

\section{Introduction}
\label{sec:introduction}

Due to the limited sensitivity achievable with conventional techniques,
polarimetry of X-ray astrophysical sources is, to date, essentially limited to high-significance
detections for a single bright source, the Crab Nebula~\cite{1978ApJ...220L.117W, 2020NatAs...4..511F}.
Gas Pixel Detectors (GPD)~\cite{Costa:2001mc} were proposed in the early 2000 as the first practical 
implementation of soft X-ray photoelectric polarimetry, with the potential for a leap in sensitivity
by more than an order of magnitude.\footnote{Previous attempts to exploit the photoelectric effect in gas were effective at higher energy, see e.g.~\cite{AustinRamsey}.}
This technology opened the way to mission concepts offering
for the first time the opportunity to observe tens of sources for precision measurements of their
polarimetric properties, providing invaluable insight into their geometries and the physical
processes at play.

After about two decades of dedicated R\&D to bring the GPD technology readiness 
level to flight standards, and a demonstration (without use of imaging property) on the PolarLight
CubeSat~\cite{2019ExA....47..225F}, 
IXPE~\cite{2019SPIE11118E..0VO} will be the first mission 
to fully exploit this technology,
starting the first ever polarimetric sky study in X-rays at the end of 2021.

In this work we describe the IXPE GPD, as matured in the last decade and verified
with a number of prototype detectors evolved towards the 
final design. We assembled and tested 9 GPDs of this design, of which we selected 4 to be incorporated into flight-model detector units (3 flight plus a spare).
We provide details of the detector key components, the front-end Application Specific Integrated Circuit (ASIC) enabling the photoelectron
track reconstruction, the readout electronics controlling the signal digitization, and
the mechanical design ensuring the necessary structural and thermal robustness.

\begin{figure}[htb!]
    \centering
    \includegraphics[width=\linewidth]{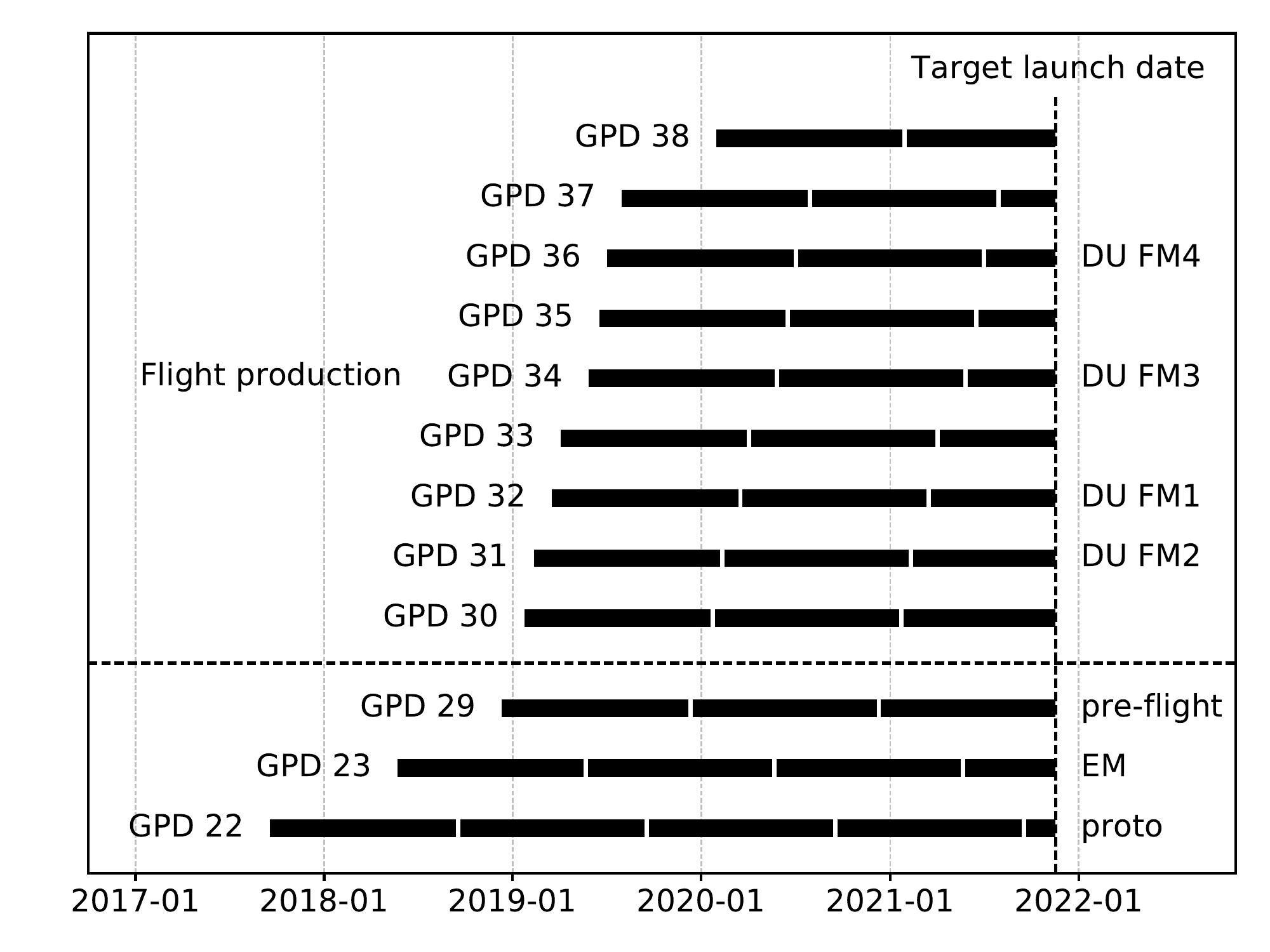}
    \caption{Temporal development of the GPD pre-flight and flight production, with the 
    black horizontal bars representing the operational period for each GPD.
    (The white vertical ticks indicate 1-year time intervals since the GPD sealing.)
    The sequential numbering scheme for the detectors is an heritage of the R\&D activity;
    Detector Units (DU) Flight Models (FM) 2--4 are currently integrated on the focal plane
    of the satellite, while DU FM 1 is the flight spare. By the launch date we shall have an integrated total of$\sim 33$~GPD-years of detector operation on the ground.}
    \label{fig:gpd_inventory}
\end{figure}

Having accumulated the equivalent of $\sim 25$~GPD-years of test data at the time of writing
(see Figure~\ref{fig:gpd_inventory}), we uncovered some more subtle instrumental effects that
are now understood and must be taken into account with dedicated calibrations during the IXPE mission. 
Minimizing these effects in an improved GPD design offers an exciting
research opportunities for the next generation of astrophysical polarimeters.

\section{Design Drivers}
\label{sec:design}

\begin{figure}[t!]
  \centering\includegraphics[width=\linewidth]{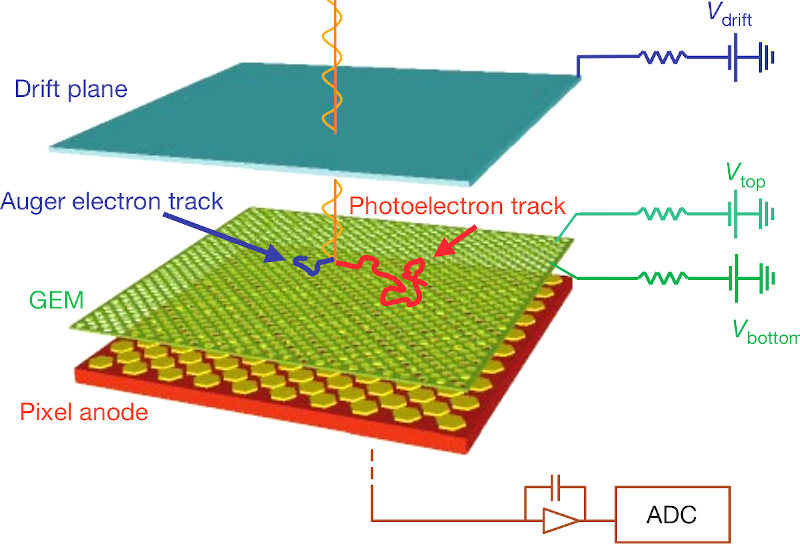}
  \caption{Conceptual design of the GPD (adapted from~\cite{Costa:2001mc}). The volume of
  the gas cell is divided into two parts: the (upper) absorption gap, between the drift plane (which is also the entrance
  window) and the GEM top, and the (lower) transfer gap, 
  between the GEM bottom and
  the readout anode plane (the readout ASIC).}
  \label{fig:gpd_concept}
\end{figure}

Figure~\ref{fig:gpd_concept} illustrates the conceptual design of the gas pixel detector,
as well as the basic detection principle. Photons enter the active gas volume through a
beryllium window, and can be absorbed in the gas. Under the action of the electric field (parallel to the optical axis)
in the absorption gap, the primary ionization electrons 
generated by the photoelectron drifts toward
the Gas Electon Multiplier (GEM)~\cite{SAULI1997531}, which provides the necessary gas gain while preserving the track shape.
Finally, the charge generated in the avalanche is collected on the readout
ASIC~\cite{BELLAZZINI2004477,BELLAZZINI2006552}, acting as a finely pixellated anode.
The polarization information is recovered on a statistical basis from the
azimuthal distribution of the photoelectron directions of emission, reconstructed
by imaging the track projections onto the readout plane.

\begin{table}[b!]
    \centering
    \begin{tabular}{p{0.6\linewidth}p{0.32\linewidth}}
    \hline
    Parameter & Value\\
    \hline
    \hline    
    Thickness of the absorption gap & 10~mm\\
    Thickness of the transfer gap & 0.7~mm\\
    Thickness of the Be window & 50~$\mu$m ($+50$~nm Al) \\
    Active area & $15 \times 15$~mm$^2$\\
    Readout pitch & 50~$\mu$m\\
    Gas Volume & $60 \times 60 \times 10$~mm$^3$\\
    Gas mixture & Pure DME\\
    Filling pressure & 800 mbar\\
    Typical $V_\text{drift}$ & $-2800$~V\\
    Typical $V_\text{top}$ & $-870$~V\\
    Typical $V_\text{bottom}$ & $-400$~V\\
    $V_\text{ASIC}$ & $\sim 0$~V\\
    Operating temperature & +15 $^{\circ}$C to +30 $^{\circ}$C\\
    \hline
    \end{tabular}
    \caption{Summary table of the basic characteristics of the Gas Pixel 
    Detectors for the IXPE mission.}
    \label{tab:gpd_characteristics}
\end{table}

The basic parameters determining the sensitivity of a polarimeter are its quantum efficiency
$\varepsilon$ and its modulation factor $\mu$---the latter representing the response to
100\% linearly polarized radiation in the form of a single number ranging from 0 (for a detector
with no polarization sensitivity) to 1 (for a perfect polarimeter).
The two are customarily combined (see, e.g., \cite{10.1117/12.857357}) into a
single figure of merit called the \emph{minimum detectable polarization} (MDP):
\begin{align}\label{eq:mdp}
    \text{MDP} \propto \frac{1}{\qfact}, \quad \text{with} \quad  \qfact = \mu \sqrt{\varepsilon}
\end{align}
where $\qfact$ is the quality factor.
Notably, the inverse of the MDP scales linearly with  the modulation factor and only as the square
root of the quantum efficiency. Much of the trade-offs that went into the design of the
IXPE gas pixel detectors are readily understood as a coherent attempt at maximizing
the quality factor $\qfact$ in the target 2--8~keV energy band.

The GPD, whose main characteristics are summarized in Table~\ref{tab:gpd_characteristics}, is one of the key ingredients in achieving the IXPE target polarization sensitivity.
In addition, as we shall see in Section~\ref{sec:performance}, it is naturally suited
for precise ($\sim 1~\mu$s) time tagging of the events, which will give access to time-resolved
polarimetry in classes of sources such as accreting pulsars and binary systems,
and provides moderate spectroscopic capabilities (at the level of $\sim 17\%$ FWHM at 5.9~keV, roughly scaling as $1/\sqrt{energy}$),
enabling energy-resolved polarimetry where statistics is large enough.
With the design focal length of 4~m, the GPD is a good match for the IXPE 
optics, providing an intrinsic spatial resolution significantly better than the half-power diameter (HPD)
of the optics over a field of view of $\sim 12$~arcmin, sufficient to cover the vast 
majority of the extended sources that we shall observe.

\subsection{The Choice of the Filling Gas}

The choice of the gas acting as the absorbing medium is a complex trade-off involving
several different aspects of detector operation and performance.
Heavy elements are favoured from the standpoint of the quantum efficiency; however,
light gas mixtures provide a favourable stopping-power/scattering ratio, which translates
into relatively longer and straighter tracks, allowing for a higher modulation factor.
For our specific application, since  the photoelectron emission direction is 100\%
modulated only for $S$ orbitals, working above the $K$-edge of the absorbing material
is critical, which further limits the maximum $Z$ that can be used. In addition, the effect
of the atomic relaxation via the (isotropically distributed) emission of an Auger electron
becomes one of the dominant limiting factors for the low-energy polarimetric sensitivity.
In practice oxygen, with a $K$-edge of 525~eV is the heaviest candidate atomic element that
can be considered in the $2$--$8$~keV energy band.

Noting that this effectively leaves out all the noble gases customarily exploited in 
traditional proportional counters, we choose pure dimethyl-ether (DME, (CH$_3$)$_2$O, see Table~\ref{tab:dme_properties}) as a
good compromise between the various design considerations. DME has a long history of
application in gas detectors for high-energy physics, and its quenching properties are
desirable in our application, as they limit the risk of accidental discharges
in the detector. In addition, DME features one of the lowest transverse diffusion coefficients,
which is also desirable, as in practice the track blurring due to diffusion is one of the
limiting factors to our ability to reconstruct the photoelectron emission direction.

\begin{table}[b!]
    \centering
    \begin{tabular}{p{0.6\linewidth}p{0.3\linewidth}}
    \hline
    Parameter & Value\\
    \hline
    \hline
    Chemical composition & (CH$_3$)$_2$O\\
    Density @ 1 atm, 0 degrees C & $2.115$~mg~cm$^{-3}$\\
    Average energy per electron/ion pair & 28 eV\\
    Fano factor & $\sim$~0.30\\
    Minimum transverse diffusion & 68~$\mu$m~cm$^{-\frac{1}{2}}$\\
    \hline
    \end{tabular}
    \caption{Summary table of the relevant properties of dimethyl-ether (DME), 
    see, e.g., \cite{dmeprop} and \cite{Sharma}.}
    \label{tab:dme_properties}
\end{table}

\subsection{Geometrical Detector Layout}

The thickness of the absorption gap is the single, most important geometrical parameter
determining the polarimetric performance of the GPD. Although a thicker gap provides a relatively
higher quantum efficiency, the corresponding increase of drift length for the primary ionization
causes a decrease of the modulation factor due to the track blurring induced by the transverse
diffusion. The  effect is somewhat exacerbated when the absorption efficiency approaches unity,
as in the optically-thick regime photons tend to convert primarily in the uppermost layer of
the absorbing medium, further increasing the average drift length.

\begin{figure}[htbp!]
    \centering
    \includegraphics[width=\linewidth]{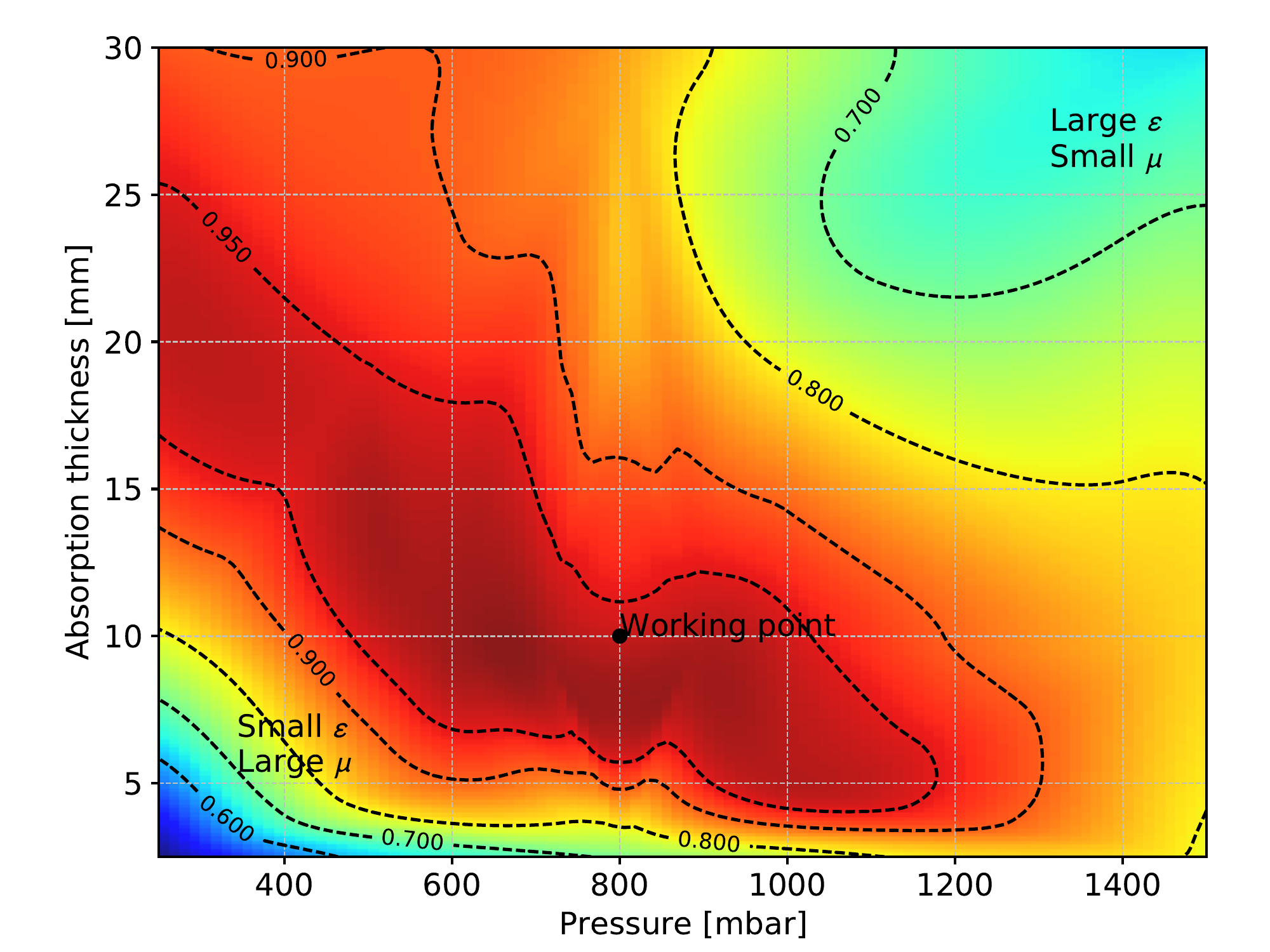}
    \caption{Relative scaling of the quality factor $\qfact$ in~\Eqref{eq:mdp}
    at 3~keV as a function of the DME pressure and the thickness of the absorption gap.
    The maximum value is conventionally set to $1$.}
    \label{fig:gpd_design_tradeoff}
\end{figure}

\begin{figure}[!htb]
    \centering
    \includegraphics[width=\linewidth]{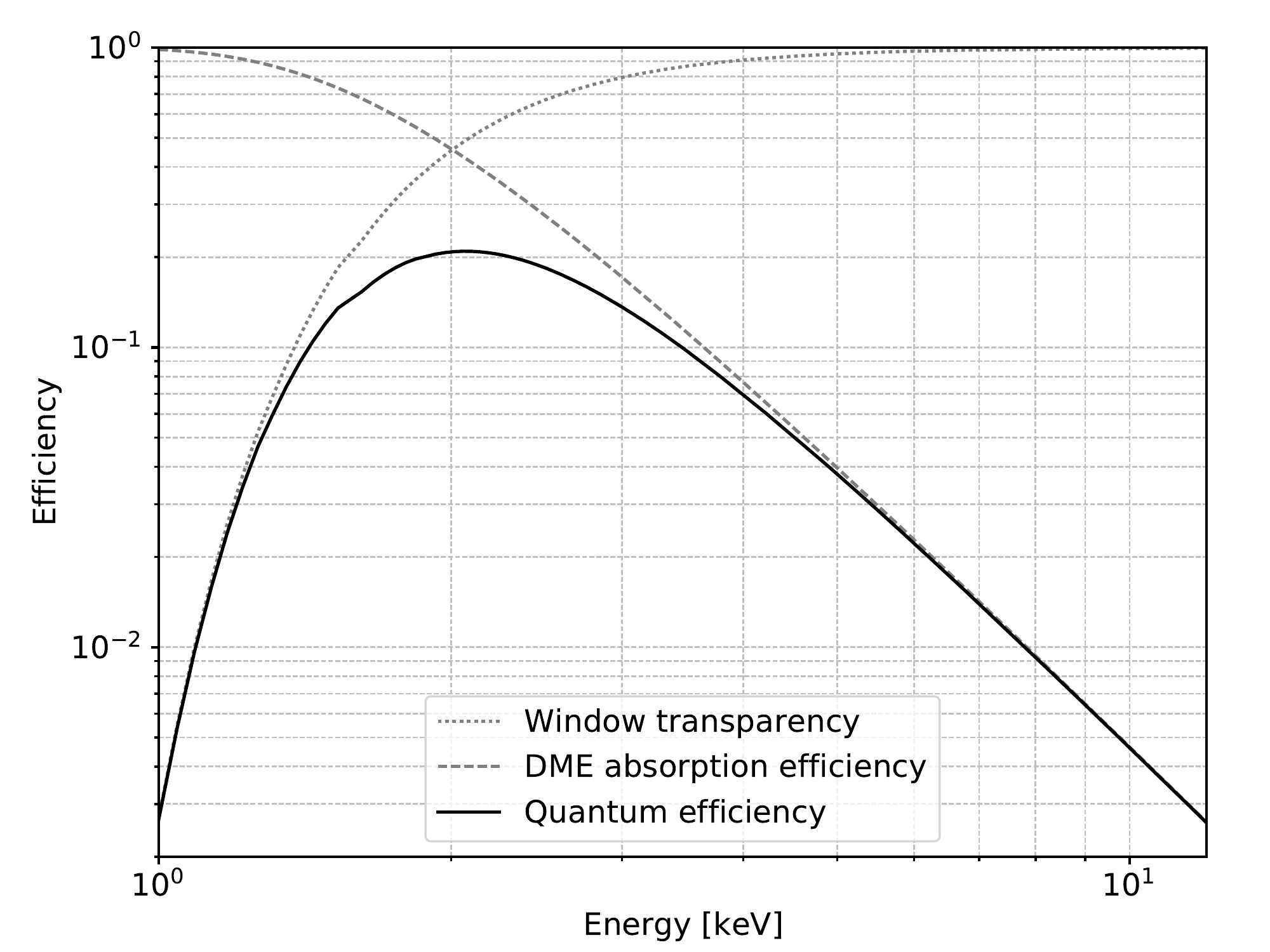}
    \caption{GPD quantum efficiency as a function of energy. The calculation assumes
    1~cm of DME at $800$~mbar and $20^\circ$~C, and includes the effect of the 
    $50$~nm Al deposition on the inner face of the Be window, as well as that of 
    the contaminants in the window itself (mainly BeO and Fe, the Be purity
    being 99.0\%).}
    \label{fig:quantum_efficiency}
\end{figure}

The optimization of the geometry of the gas cell is tightly coupled to the choice of the gas
pressure, and Figure~\ref{fig:gpd_design_tradeoff} shows the  quality factor $\qfact$
in~\Eqref{eq:mdp}, calculated at 3~keV, in the absorption thickness-pressure plane%
\footnote{
The actual broadband sensitivity depends on the effective area of the X-ray optics and the
source spectrum, but, being 3~keV close to the energy of peak sensitivity, this is a sensible
proxy, providing a good illustration of the expected performance across the phase space of interest.}.
Although we emphasize that our sweet spot is fairly shallow, 10~mm of pure DME at 800~mbar
provide a nearly optimal sensitivity in a configuration that is well matched to basic detector
components: with an average photoelectron track length ranging from about $100~\mu$m at 2~keV
to slightly over $1$~mm at $8$~keV, the $50$~$\mu$m pitch of the readout ASIC and of the GEM
allows a meaningful reconstruction of the track morphology across the entire energy band,
and is comparable to the characteristic scale of the transverse diffusion. When convolved with
the transparency of the entrance window, this geometrical arrangement provides an overall peak
quantum efficiency in excess of $20\%$ at $2$~keV, dropping to about $1\%$ at $8$~keV, as shown
in~Figure~\ref{fig:quantum_efficiency}.

The transfer gap acts mainly as a physical separation between the bottom face of the
GEM and the readout plane and, as such its thickness should be in principle as small as
possible to avoid additional track blurring. The nominal value of $700~\mu$m that we choose
is mainly driven by the distance between the bottom face of the GEM and the top of the
wire-bonding loops from the ASIC to the ceramic package hosting it.

We also emphasize that the footprint of the gas cell ($4 \times 4$~cm$^2$) is significantly
larger than the active area of the readout chip, which has the twofold benefit of 
guaranteeing a more uniform electric field in the absorption gap and reducing the effect of
possible background generated in the ceramic walls of the cell itself~\cite{10.1117/12.925385}.

\section{The Gas Electron Multiplier}
\label{sec:gem}

The Gas Electron Multiplier~\cite{SAULI1997531} provides the gain stage for the GPD.
Being intrinsically two-dimensional, it is particularly suited for our application.
Compared with other GEM devices customarily used in high-energy physics applications,
the main peculiarity of those developed for the IXPE mission is their fine pitch,
which is in turn dictated by the necessity to preserve as much as possible the morphology
of the photoelectron track and match the sampling capabilities of the readout plane.

We choose the laser-etching technique described in~\cite{TAMAGAWA2006418} as a well proven
technology for producing a GEM with such a small pitch. As a matter of fact, the production process
was fine-tuned through the development phase of the mission, pushing the manufacturing technology
to the limits.

\begin{table}[t!]
    \centering
    \begin{tabular}{p{0.42\linewidth}p{0.48\linewidth}}
    \hline
    Parameter &	Value\\
    \hline
    \hline
    Number of holes & $112008~(359 \times 312)$\\
    Horizontal pitch & $43.30$~$\mu$m\\
    Vertical pitch & $50.00$~$\mu$m\\
    Hole diameter & $30$~$\mu$m\\
    Hole diameter dispersion & $\sim 1$~$\mu$m (typical)\\
    Top-bottom alignment & $\sim 2$~$\mu$m (typical)\\
    Metal coating & Copper\\
    Coating thickness & $5$~$\mu$m\\
    Substrate & Liquid crystal polymer (LCP)\\
    Substrate thickness & $50$~$\mu$m\\
    Manufacturing process & Laser etching\\
    Typical operating voltage & $\sim 470$~V\\
    Gain gain scaling & $\propto \exp(\sim 0.03$~V)\\
    Working effective gain & $\sim 200$\\
    \hline
    \end{tabular}
    \caption{Summary table of the basic GEM characteristics.
    The gain scaling is a single parameter expressing the approximate 
    gain characteristics in the purely exponential regime, and represents
    the fractional gain increase per unit voltage increase ($\sim 3\%$~per~V for the 
    IXPE GEMs).}
    \label{tab:gem_characteristics}
\end{table}

\begin{figure}[b!]
    \centering
    \includegraphics[width=0.9\linewidth]{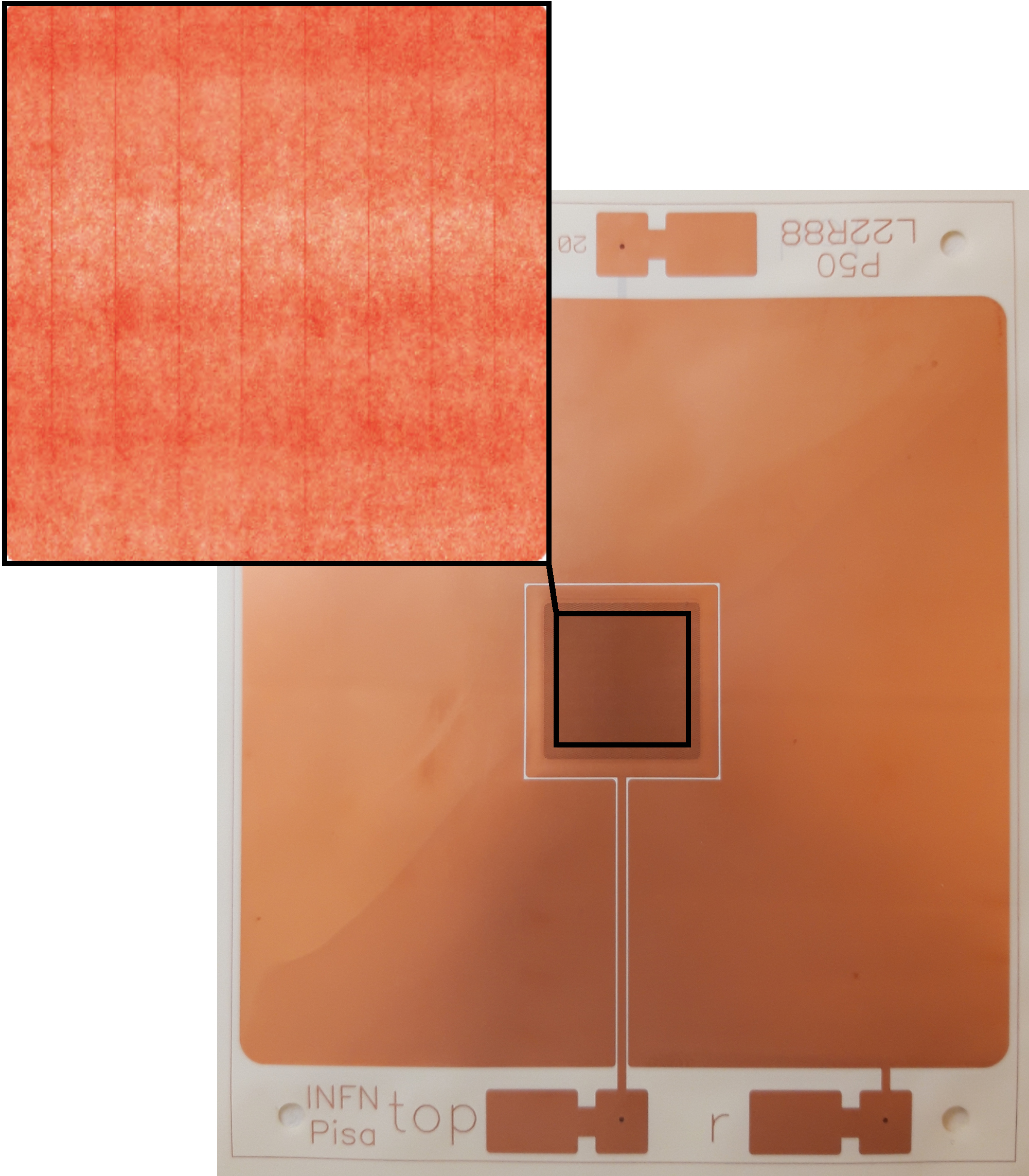}
    \caption{Photograph of a flight model GEM. The active area is the darker region in the
    center, while the rest of the copper is the guard ring on the top surface.
    The soldering pads for the high-voltage cables are visible on the very top and the
    very bottom of the image. The micro-photograph of the GEM active region shows the
    (vertical) features in correspondence to the sweep overlaps at the laser
    drilling stage, as explained in the main text. We emphasize that the second pass of
    the laser etching, in the orthogonal direction, is performed from the opposite side
    of the GEM, and therefore the horizontal features are not visible on the top surface
    pictured in the inset.}
    \label{fig:gem}
\end{figure}

The main characteristics of the GEM are summarized in Table~\ref{tab:gem_characteristics},
and a picture of a flight model is shown in~Figure~\ref{fig:gem}. 
The hole pattern follows a hexagonal grid matching the pitch of the ASIC.
The active area of the GEM is $\sim 0.5$~mm larger, on all four sides, than that of the
readout ASIC, to compensate for a possible misalignment in the assembly. 
A guard ring of about $3.5$~cm surrounding the top face of the GEM, matching the footprint
of the drift electrode, helps improving the uniformity of the drift field
and reducing possible edge effects.

\subsection{GEM Manufacturing}\label{sec:gem_manufacturing}

Although the details of the GEM manufacturing is not the primary focus of this paper,
the matter is relevant for the discussion of the systematic effects in~Section~\ref{sec:systematics}.
The GEMs are produced by SciEnergy in Japan in collaboration with RIKEN.
Roughly speaking, the manufacturing process can be broken up into three main steps:
(i) the holes are drilled in the top and bottom copper layers through standard chemical etching;
(ii) the foil is irradiated with a de-focused laser to drill the holes into the dielectric
substrate, with the residual copper acting as a mask;
(iii) a wet-etching post-processing is applied in order to polish the copper
surfaces and ensure the necessary robustness against micro-discharges.

Among the three manufacturing steps, the second is noteworthy in that it leaves a definite
geometrical imprinting in the GEM foils. The laser drilling of the substrate is achieved by
using a $1.8$~mm-wide laser beam sweeping the GEM active surface multiple times to cover
the entire area, with a small overlap of about $100~\mu$m between successive passes.
In order to reduce the amount of heat that needs to be dissipated in the process, the holes are
drilled halfway through from the top GEM surface, with sweeps in the vertical direction, and
completed from the bottom surface with sweeps in the horizontal direction. The net result is that
even a naked-eye optical inspection of the GEM reveals 8 thin horizontal stripes and 8 thin
vertical stripes (spaced by 1.8~mm) at the overlap positions of adjacent laser sweeps, as shown
in the inset in Figure~\ref{fig:gem}.

At a microscopic level this affects the properties of the holes in a way that,
as we shall see in Section~\ref{sec:systematics}, has important implications for the
polarimetric response of the detector.

\subsection{GEM Screening and Functional Tests}

Each finished GEM foil undergoes a thorough optical metrology to assess the
accuracy in the mask alignment, the overall quality of the holes across the active
surface and the possible presence of visible defects. This provides valuable information
for the selection of the foils to be used for flight detectors.
The initial screening proceeds with basic electrical tests: a verification of the 
isolation between the top and the bottom electrodes (up to 250~V in air) and a measurement
of capacitance of the GEM and the ring.

Most of the functional tests prior to the integration of the actual detectors are
performed in a \emph{test-box} setup equipped with a readout ASIC identical to those
used for flight and with a series of spacers specifically designed to hold the GEM
in place and apply the high-voltages on the proper soldering pads by mechanical
pressure (i.e., without the need to solder the cables on the pads themselves).
The box is fluxed with a mixture of Ar/CO$_{2}$ 70/30 at atmospheric pressure, and 
a thin window acting as the drift electrode allows us to test of the assembly with a
radioactive $^{55}$Fe source, in a geometrical configuration similar to the flight 
detectors.
While the gas mixture used in the test box is not suitable for measuring the 
polarimetric response, it is useful to measure the energy resolution
and the gain uniformity of the GEM with X-rays, as well as the possible presence of localized defects 
(e.g., dead or hot spots) to be correlated with the outcome of the optical inspection.
(We emphasize, however, that the quality of the batches used for the GPD 
flight production was outstanding, with little or no presence of such defects.)

\begin{figure}[htbp!]
  \centering\includegraphics[width=\linewidth]{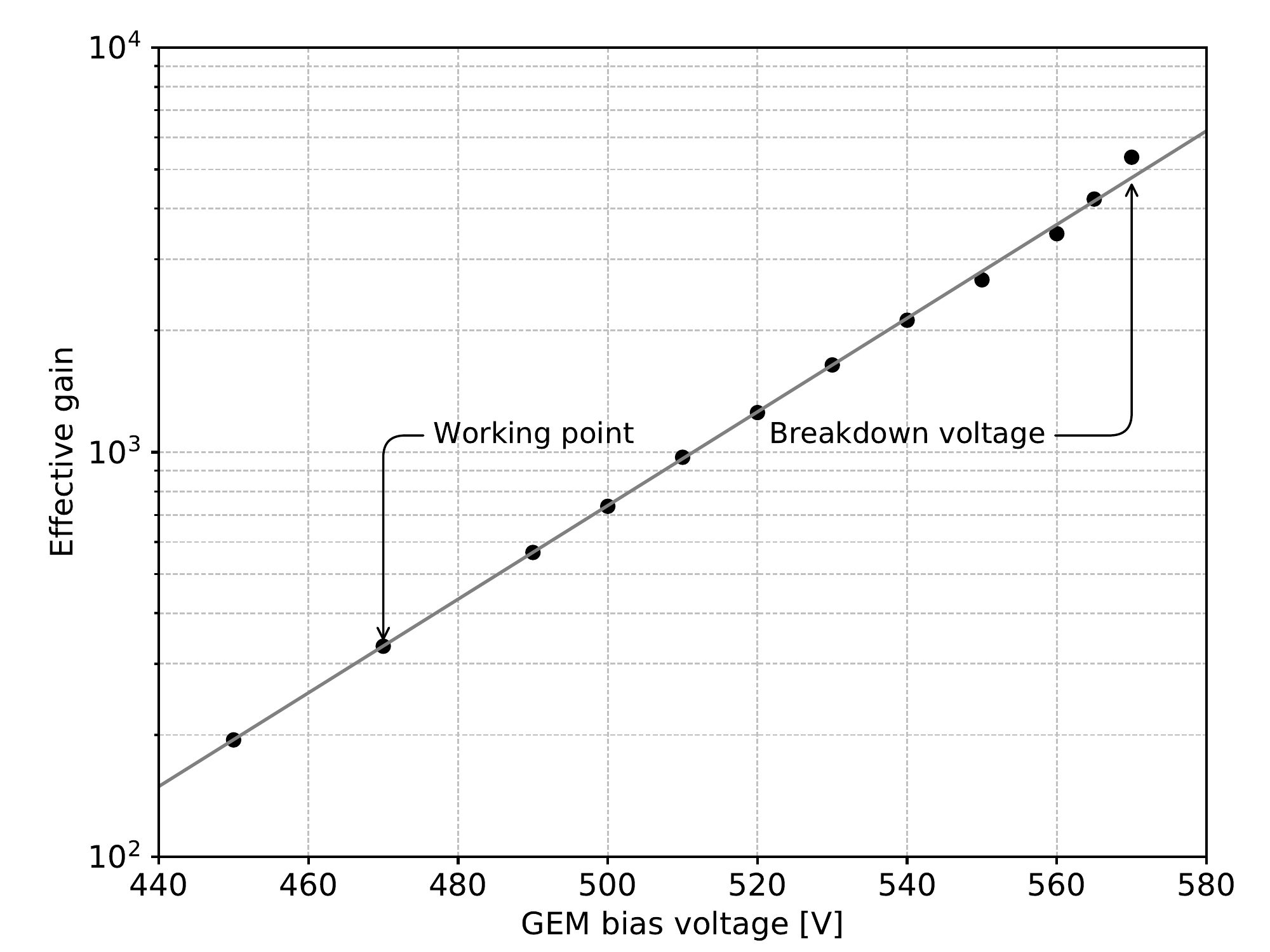}
  \caption{Typical GEM gain-voltage characteristics in~Ar/CO$_{2}$ 70/30 at 1~bar.
  The breakdown voltage corresponds to an effective gain $\sim 15$ times
  higher than our nominal working point.}
  \label{fig:gem_gain_characteristics}
\end{figure}

In addition to the standard tests that all the GEM foils undergo, we do perform sample
destructive measurements on at least one foil per batch in a test setup with a single
readout pad connected to a charge amplifier.
Figure~\ref{fig:gem_gain_characteristics} shows a typical GEM gain characteristics
in~Ar/CO$_{2}$ 70/30 at 1~bar, increasing exponentially at a rate of $\sim 3\%$ per V 
(i.e., doubling every $\sim 25$~V). Due to the low readout noise, we typically 
operate at a gas gain of a few hundreds, and the breakdown voltage is
$\sim 100$~V above our working point, at an effective GEM gain more than
$10$~times larger.
The excellent uniformity of the gain characteristics, with the index of the 
exponential typically varying by $\pm 5\%$ across different GEM foils, provides
evidence that we will safely operate in space, far from the discharge regime%
\footnote{Strictly speaking these figures are measured with a different gas mixture
with respect to the one used in flight, but the gain characteristics in pure DME at
800~mbar are very similar---modulo a slightly lower normalization, roughly
corresponding to a difference of $10$~V in bias voltage. In addition, the quenching 
properties of the DME provide additional robustness against micro-discharges.}%
.

\section{The Readout ASIC}
\label{sec:asic}

\begin{figure}[htbp!]
  \centering\includegraphics[width=\linewidth]{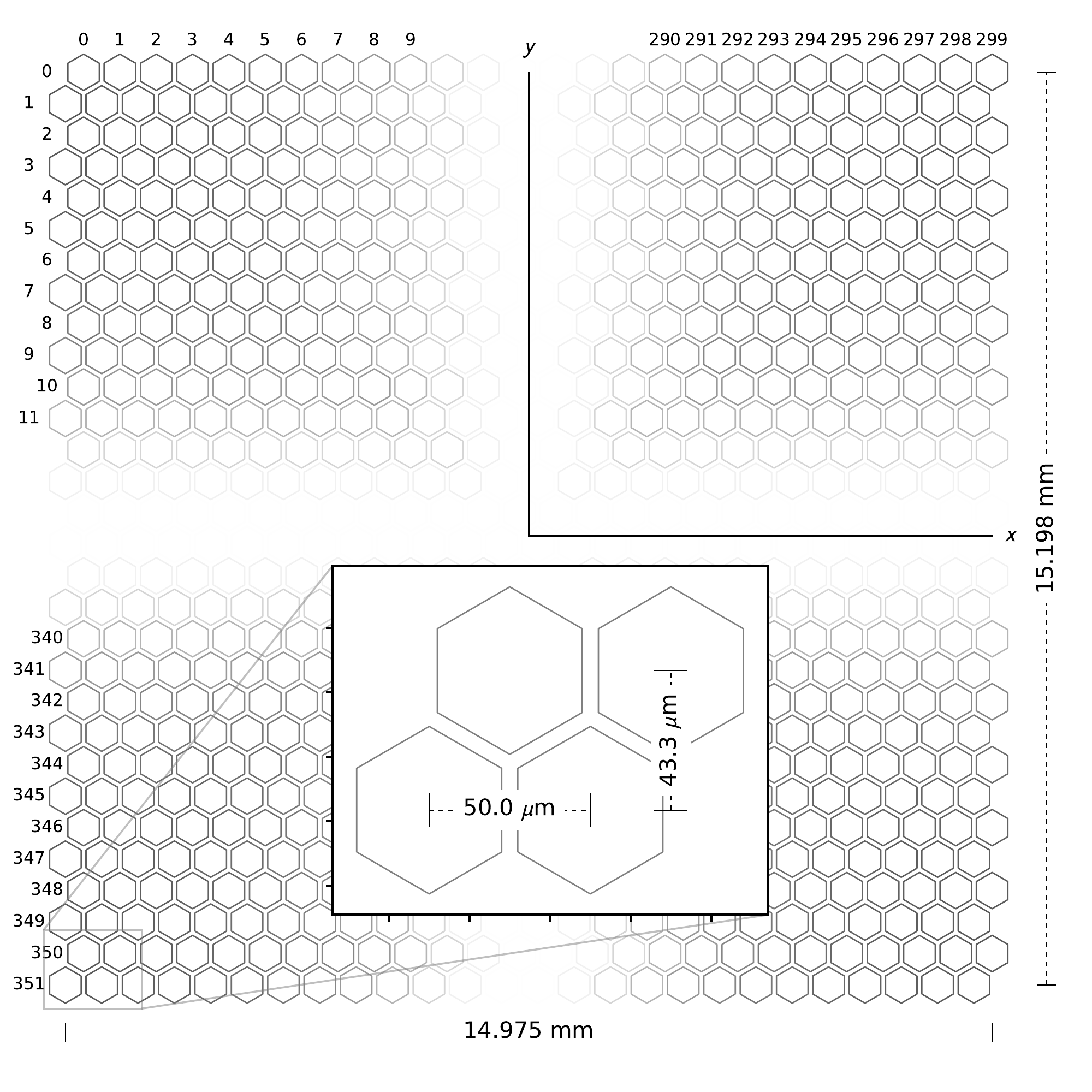}
  \caption{Geometrical layout of the readout ASIC. The inset illustrates the 
  structure of a trigger mini-cluster, which is further described in 
  Section~\ref{subsec:trigger}.}
  \label{fig:xpol_geometry}
\end{figure}

The ASIC~\cite{BELLAZZINI2004477,BELLAZZINI2006552} acts as a readout anode for the GPD.
The chip, based on 0.18 $\mu$m CMOS technology, integrates more than 16.5 million transistors and is organized as a
matrix of 105,600~pixels (300 columns at $50.00$~$\mu$m pitch and 352 rows at $43.30$~$\mu$m
pitch) with a $15 \times 15$~mm$^2$ active area, as shown in Figure~\ref{fig:xpol_geometry} and Table~\ref{tab:asic_characteristics}.

\begin{table}[htbp!]
    \centering
    \begin{tabular}{p{0.5\linewidth}p{0.4\linewidth}}
    \hline
    Parameter &	Value\\
    \hline
    \hline
    Number of pixels & $105600~(300 \times 352)$\\
    Horizontal pitch & 50.00~$\mu$m\\
    Vertical pitch & 43.30~$\mu$m\\
    Shaping time & 4~$\mu$s\\
    Pixel gain & $\sim 400$~mV~fC$^{-1}$\\
    Pixel Noise & $22.5$~$e^{-}$~ENC\\
    Dynamic range & 1~V ($\sim 30\rm{k}~e^{-}$)\\    
    \hline
    \end{tabular}
    \caption{Summary table of the basic readout ASIC characteristics.}
    \label{tab:asic_characteristics}
\end{table}

Each pixel is composed of a hexagonal metal electrode connected to a charge-sensitive
amplifier followed by a shaping circuit. The ASIC provides a built-in, customizable
self-triggering capability, complemented with an on-chip signal processing for automatic
localization of the event. Upon trigger, the maximum of the shaped pulse is stored inside
each pixel cell for subsequent readout.
Additionally, the chip features an internal charge-injection system that can be used
to  stimulate any pixel in the matrix, for diagnostic and calibration purposes.

\subsection{Back-End Electronics}

The Back-End Electronics (BEE) is responsible for all the commanding and control of
the readout chip, as well as the generation of low- and high-voltages for the detector
and the handling of the science data and telemetry interfaces~\cite{9405651}.
At the hardware level, the BEE consists of three distinct electronic boards plugged
onto a common backplane: a low-voltage power supply, a high-voltage power supply
and a data acquisition (DAQ) board. (For completeness, all the tests described in this paper have
been performed with a commercial HV power supply, 
since models with flight-like design were not yet available for this activity.)

The DAQ board incorporates two fundamental components of the GPD operation, i.e., the
analog-to-digital converter for the serial readout and the Field Programmable Gate Array (FPGA) controlling the ASIC
configuration, control and readout, as well as the data formatting and the communication with
the detector service unit.
The ADC on the DAQ board has a resolution of 14 bits over a 2.4 V full dynamic range,
or a voltage resolution of 0.146 mV/ADC count. Coupled to the nominal pixel gain,
this translates into a charge characteristics for the whole system of 2.3 electrons/ADC count. The $1$~V dynamic range of the pixel amplifiers in the readout ASIC 
results in a saturation value of about $6500$~ADC counts, 
more than 5 times larger than the typical
pixel signal in the Bragg peak at the nominal GEM gain.
Table~\ref{tab:bee_summary} shows a summary of the basic characteristics of the readout electronics.

\begin{table}[htbp!]
    \centering
    \begin{tabular}{p{0.5\linewidth}p{0.4\linewidth}}
    \hline
    Parameter & Value\\
    \hline
    \hline
    Dynamic range & $2.4$~V ($-1.2$~V to $1.2$~V)\\
    ADC Resolution & $14$~bit\\
    ADC voltage resolution & $0.146$~mV~fC$^{-1}$\\
    Charge characteristics & $2.3~e^-$ per ADC count\\
    Typical gain & $3000$~ADC counts~keV$^{-1}$\\
    \hline
    \end{tabular}
    \caption{Basic characteristics of the readout electronics used for the GPD tests.
    Note that the conversion factor from keV to ADC counts quoted in the table
    is largely driven by the GEM gain.}
    \label{tab:bee_summary}
\end{table}

\subsection{Event Triggering}
\label{subsec:trigger}

At the very fundamental level every 4 adjacent pixels (see the inset in
Figure~\ref{fig:xpol_geometry}) are logically OR-ed together to contribute to a
local trigger with a dedicated, fast shaping amplifier.
This basic building block of $2 \times 2$ pixels is called a trigger
\emph{mini-cluster}, and is central to the entire machinery of event triggering
and readout.

Upon trigger, the event is automatically localized by the ASIC in a rectangle
containing all triggered mini-clusters plus a padding of 4 or 5 additional
ones%
\footnote{Since, in normal operating conditions, only the part of the track with
the  largest specific ionization (i.e., the Bragg peak) participates into the trigger,
the padding serves the fundamental purpose of capturing to the readout the
entire track---including the parts with a relatively low ionization density.}
along the $X$ and $Y$ coordinates, respectively, in order to 
compensate as much as possible for the uneven aspect ratio of the mini-cluster.
More specifically, the chip calculates the coordinates of the upper-left and
lower-right corners of such a rectangular region of interest (ROI), 
and limits the serial readout to the subset of pixels 
(typically 500--800) within that ROI.
This allows for a reduction of the readout time of more than 
two orders of magnitude, compared to that for a complete frame.

\subsection{Event Readout and Pedestal Subtraction}

Upon definition of the region of interest, the serial readout proceeds driven
by a dedicated readout clock, generated by the back-end electronics. The analog
output of each pixel is sequentially routed to the differential output buffer
of the ASIC, which in turn is connected to the ADC on the DAQ board.
The typical settling time of the output buffer is of the order of $\sim 200$~ns,
which limits to about 5~MHz the maximum clock frequency that can be used for serial
readout without introducing potential readout artifacts.

During nominal data taking the region of interest corresponding to each physical
track is read out two times in close succession. The readout following the 
first one is used by the DAQ board to perform a pedestal subtraction,
and the resulting stream is zero-suppressed and compressed prior to being passed
downstream. At the cost of a larger dead time, the event-by-event pedestal
subtraction largely mitigates a class of potential subtle systematic effects that
we shall discuss in more detail in Section~\ref{sec:systematics_spuriousmod}.
In addition, this readout strategy compensates for any potential dependence of
the pedestals on the environmental conditions (e.g., the temperature).
\section{GPD Assembly}

\begin{figure}[htb!]
  \centering\includegraphics[width=0.72\linewidth]{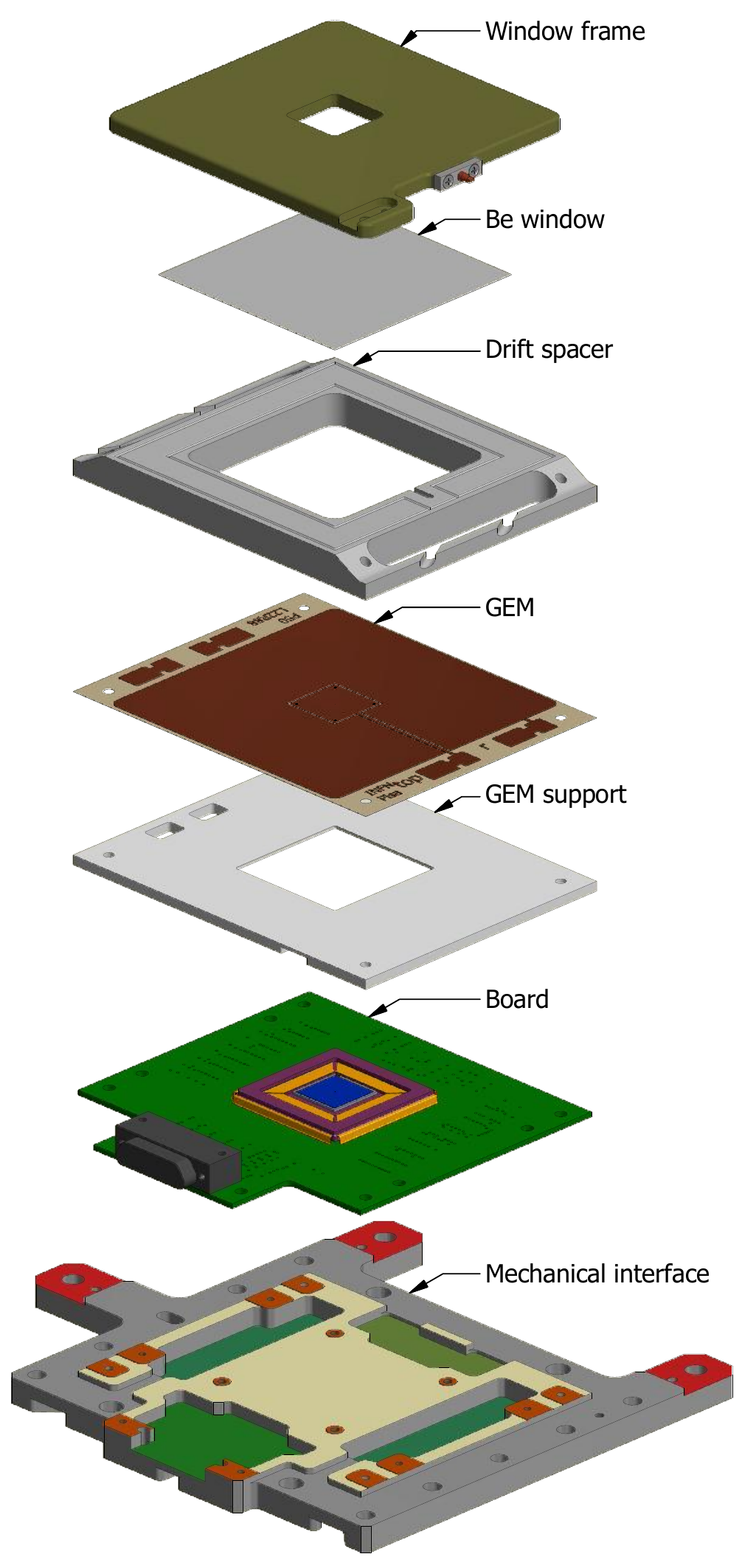}
  \caption{Exploded view of the GPD. The colors on the mechanical interface drawing identifies planes at the same height.}
  \label{fig:gpd_exploded}
\end{figure}

Figure~\ref{fig:gpd_exploded} shows an exploded view of the basic GPD elements.
The first and foremost challenge in the GPD assembly is for the stack of the
various components, made of different materials and with different thermal expansion
coefficients, to be able to sustain several thermal cycles for the glue to cure,
and a final bake-out at high temperature, and yet guarantee a leak tightness better
than $10^{-9}$~mbar~l$^{-1}$~s$^{-1}$, which is necessary for the sealed gas cell
to operate throughout the duration of the mission.
In addition, this must be accomplished within a tolerance envelope tight enough 
to allow for co-alignment of the detector active surface and the mirrors, once
the Detector Units are integrated in the satellite.
Finally, the tight out-gassing requirements severely limit the choices of the materials and the
adhesives that can be practically used.

The GPD assembly---including the wire-bonding of the readout ASIC in its package and
the positioning of the latter on the GPD board, the assembly of the gas cell, the
metrological verifications and the initial leak test---was entirely performed in house,
using the INFN facilities. The assembly procedure was developed in collaboration with our
historical industrial partner, Oxford Instruments Technologies Oy in Finland (which also performed the final bake-out and
filling of the detectors) and further refined through phases A and B of the mission.

\subsection{GPD Board Assembly}

The lowermost element of the GPD stack is a custom-design titanium frame, acting as a support
for the structure and as a mechanical reference for the entire assembly process. 
The titanium frame hosts all the elements to control the GPD temperature (two heaters on the
top face, two heaters, temperature sensors and a Peltier cell on the bottom one) and is the main thermal path for dissipating
the heat from the readout ASIC. It is also the key element for the alignment of the detectors
and the associated X-ray optics, thanks to three dedicated fingers (shown in red in
Figure~\ref{fig:gpd_exploded}) that remain accessible throughout the satellite integration.

\begin{figure}[htb!]
  \centering\includegraphics[width=0.9\linewidth]{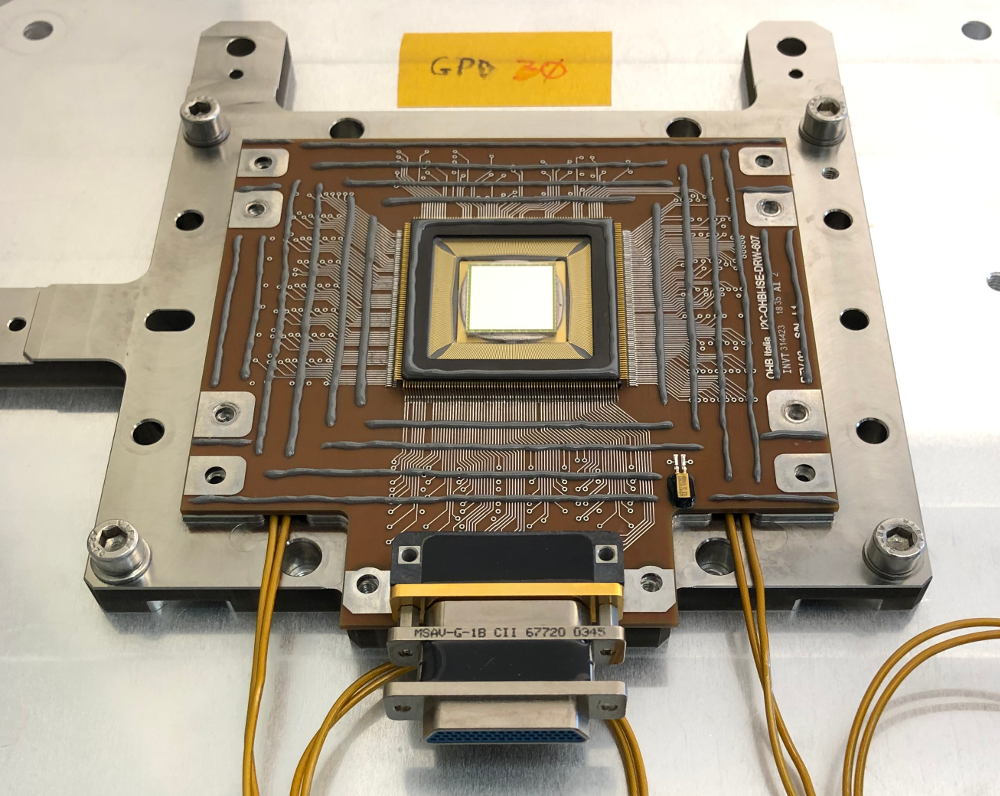}
  \caption{A GPD board on its mechanical interface, ready for the integration of the 
  GEM support, with the readout ASIC visible at the center of the picture.
  (The flight GPD boards were fabricated by OHB-I, which also performed the soldering of
  ceramic package.)
  The continuous ring of glue on top of the ceramic package is the key element
  for sealing the gas cell. The electrical cables coming out from underneath the PCB
  connect to the two heaters used for the GPD thermal control.}
  \label{fig:gpd_gpdboard}
\end{figure}

The readout ASIC is bonded with conductive glue on a commercial ceramic package, using 
a custom mechanical tool to control its positioning. The chip is then wire-bonded to the pads on
the package using a standard wedge-bonding technique, and the package is glued and soldered
on its printed circuit board (PCB). Finally, the PCB is glued on top of the mechanical interface,
as shown in Figure~\ref{fig:gpd_gpdboard}, using positioning pins for controlling the
alignment and maintaining it while the glue cures at high temperature.

The mechanical precision of this partial assembly is critical to control the thickness of
the GPD transfer gap, and 
to guarantee that the active area of the detector and its optics can be properly co-aligned.
As such, it is verified via optical metrology on a detector by detector
basis. The precision in the positioning of the ASIC is at the level of 
$\sim 50~\mu$m in the vertical direction, and $\sim 20~\mu$m in the detector plane.

\subsection{Gas Cell Assembly}

The assembly of the gas cell can be divided in two parts: we first mount
the GEM on its support frame and the beryllium window on the titanium drift frame, and then
we stack these two partial assemblies on top of the GPD board.

\begin{figure}[htb!]
  \centering\includegraphics[width=\linewidth]{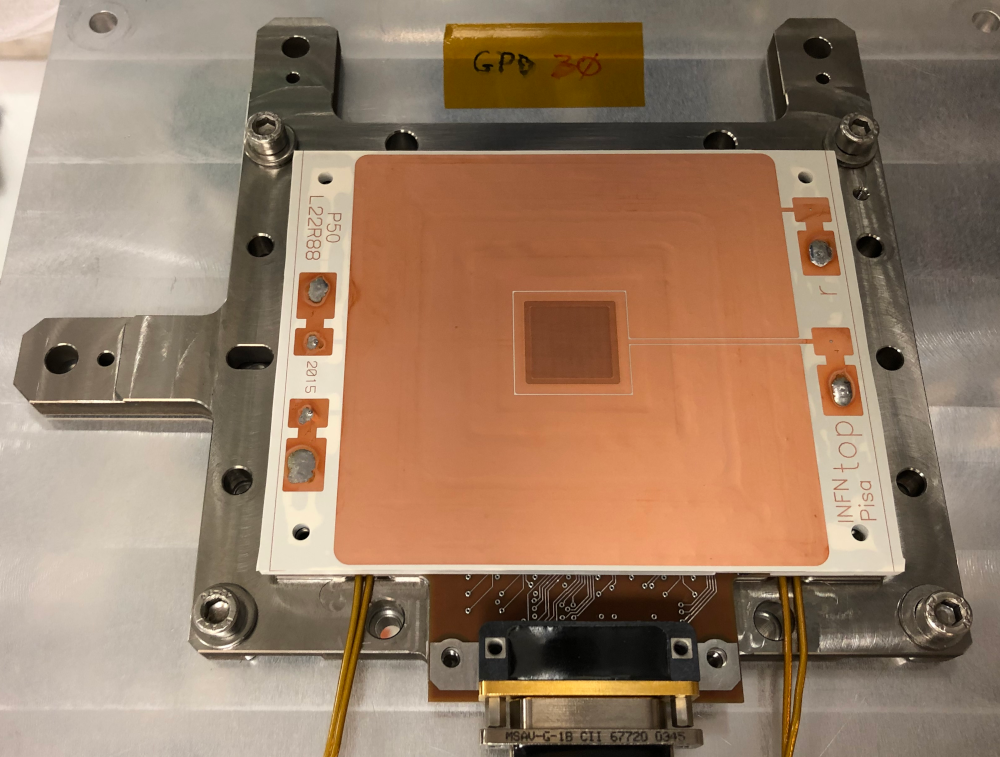}
  \caption{Picture of the GEM-Shapal spacer assembly glued onto the GPD board. This 
  partial assembly is ready for hosting the Macor spacer defing the absorption gap.}
  \label{fig:gpd_gem}
\end{figure}

The GEM is bonded to its ceramic (Shapal) support by applying a uniform tension by means
of a dedicated tool holding the GEM on a circular frame like the skin of a drum.
The GEM support is then glued directly on the PCB using alignment pins, as shown in
Figure~\ref{fig:gpd_gem}. We emphasize this is one of the crucial steps of the assembly,
as the gluing between the GEM bottom and the ASIC package 
needs to guarantee a complete leak-tightness with no possibility of further intervention.

The top sub-assembly is prepared in parallel, gluing the entrance window to its titanium frame.
The window is made from a $50~\rm{\mu m}$-thick, optical grade, high-purity beryllium foil, 
and a thin ($\sim 50$~nm) aluminum layer is sputtered on the inner face to enhance the
leak-tightness. The electrical contact between the window and the frame is guaranteed by
means of conductive glue, and output of the drift channel of the high-voltage power supply is
connected directly to the titanium frame. The latter is also holding the small copper tube
used for the gas filling.

At this point the drift titanium frame is glued to a 1~cm-thick ceramic (Macor) spacer,
defining the X-ray absorption gap, and the result is glued 
to the top of the GEM, completing the stack.
After the mechanical assembly is completed, 
we measure the final leak rate to confirm
it is below our requirement of 
$10^{-9}$~mbar~l$^{-1}$~s$^{-1}$.

\subsection{Bake-out and gas filling}

The bake-out, filling and sealing of the detectors is performed by 
Oxford Instruments Technologies Oy in Espoo, Finland.
The GPD is first connected to a vacuum system and placed in a temperature-controlled
chamber at 100~$^{\circ}$C for at least 14~days. A secondary pump is connected to a flange
placed on top of the entrance window in order to reduce the differential pressure on the 
window itself during the bake-out cycle.

\begin{figure}[htb!]
  \centering\includegraphics[width=\linewidth]{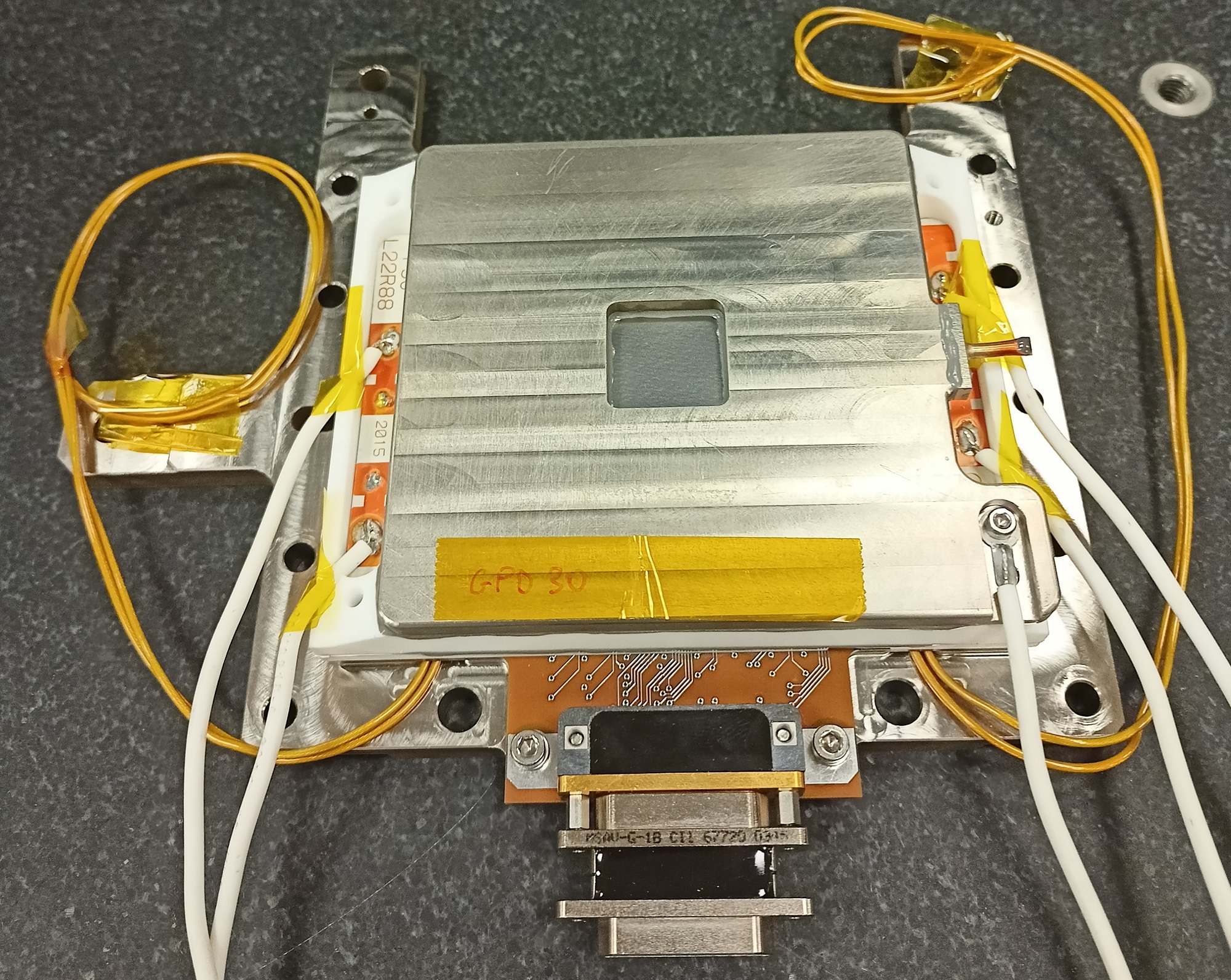}
  \caption{Picture of a complete GPD flight model ready for functional and acceptance test.
  The white electrical cable soldered on the GEM pads provide the connections to the 
  high-voltage power supply.}
  \label{fig:gpd_photo}
\end{figure}

Once the bake-out process is completed, the detector is brought back at room temperature, tested for 
leak-tightness one final time to exclude possible damage during the bake-out, and filled with
purified DME at the desired pressure (i.e., the equivalent of 800~mbar at 20~$^{\circ}$C).
After the filling tube is crimped to the final length, the complete detector, as shown in 
Figure~\ref{fig:gpd_photo}, is equipped with high-voltage cables and ready for 
functional tests.

\section{Detector Characterization}
\label{sec:performance}

We produced 9 flight GPD---out of which four were chosen to be installed in the (three plus one spare)
Detector Units (DU). 
All of them were extensively tested to verify their basic performance as focal-plane detectors (see Table~\ref{tab:gpd+performance}) using a dedicated test setup at INFN and the calibration facility
at IAPS later used for the calibration of the DUs. 
We found all the detectors to
show very similar performance metrics, as we shall detail in the remainder of this section.

\begin{table}[htbp!]
    \centering
    \begin{tabular}{p{0.4\linewidth}p{0.5\linewidth}}
    \hline
    Parameter &	Typical value\\
    \hline
    \hline
    Effective noise & 22.5 electrons ENC \\
    Gain uniformity & $\sim 20\%$ \\
    Energy resolution & $\sim 17.5\%$ FWHM at 5.9~keV \\
    Position resolution & $< 100~\mu$m rms\\
    Modulation factor & $\sim 28\%-55$\% @ 2.7--6.4 keV\\
    Spurious modulation & $<0.5\%$ at 5.9~keV\\
    Trigger efficiency & $\sim 100$\% down to 1~keV\\
    Dead time per event & $\sim 1$~ms at 2.7 keV\\
    \hline
    \end{tabular}
    \caption{Summary table of the GPD performance as a focal-plane detector.
    The reader is referred to Section~\ref{sec:systematics_spuriousmod} for a 
    detailed description of the low-energy detector response to un-polarized
    radiation.}
    \label{tab:gpd+performance}
\end{table}

\subsection{Event Reconstruction}

\begin{figure}[htbp!]
    \centering
    \includegraphics[width=\linewidth]{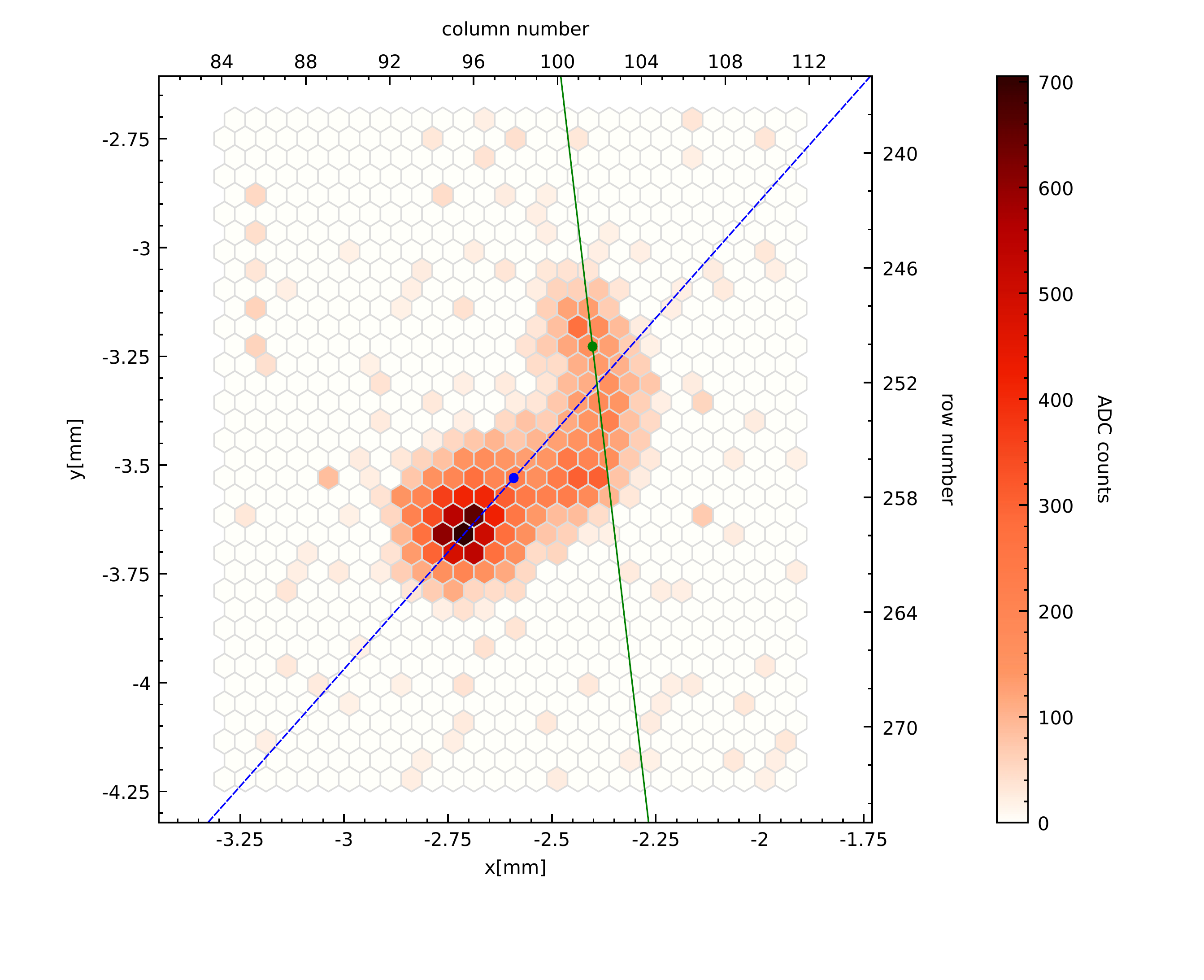}
    \caption{Example of a real track from a 5.9~keV photon, as imaged by the GPD.
    The color scale represents the charge content of each pixel, with a zero-suppression 
    threshold of 20~ADC counts (or $\sim 46$~electrons of equivalent signal charge) being 
    applied. The blue line and point represent the principal axis and the barycenter of the track,
    while the green line and point represent our best estimate of the photoelectron direction
    and photon absorption point.}
    \label{fig:sample_evt_raw}
\end{figure}

Since the event-level analysis plays an active role in determining the detector 
performance, we start this section by illustrating the basic steps involved 
in the processing of the track images, such as that shown in Figure~\ref{fig:sample_evt_raw}.
The track reconstruction starts with a zero suppression, followed by a clustering
stage (based on DBSCAN~\cite{Ester96adensity-based}) aimed at separating the physical
photoelectron track from the residual noise pixels. The reconstruction proceeds with a 
moment analysis around the barycenter to identify the principal axis of the two-dimensional
charge distribution. As most of the energy is deposited in the Bragg peak at
the end of the photoelectron path, the longitudinal charge profile can be used
to discriminate the track head from its tail and estimate the photon absorption point.
Finally, a second moment analysis is run, de-weighting the pixels close to the end
of the track, to get a more accurate estimate of the photoelectron emission direction.
At the very basic level, the reconstruction provides estimates of the event energy,
photon absorption point, photoelectron direction emission, as well as a series of
topological variables characterizing both the region of interest and the physical track, e.g., 
the size, the longitudinal and lateral extension and the charge asymmetry.

The reader is referred to~\cite{10.1117/12.459381} for more details on the standard
analysis of track images. We also point out that recent developments in this area
based on  machine-learning techniques~\cite{Kitaguchi2019ACN,PEIRSON2021164740},
while providing significant improvements in polarimetric sensitivity, are
not directly relevant for the lower-level detector performance that is the main
focus of this work.

\subsection{Noise, Gain, and Uniformity of Response}

\begin{figure}[htbp!]
    \centering
    \includegraphics[width=\linewidth]{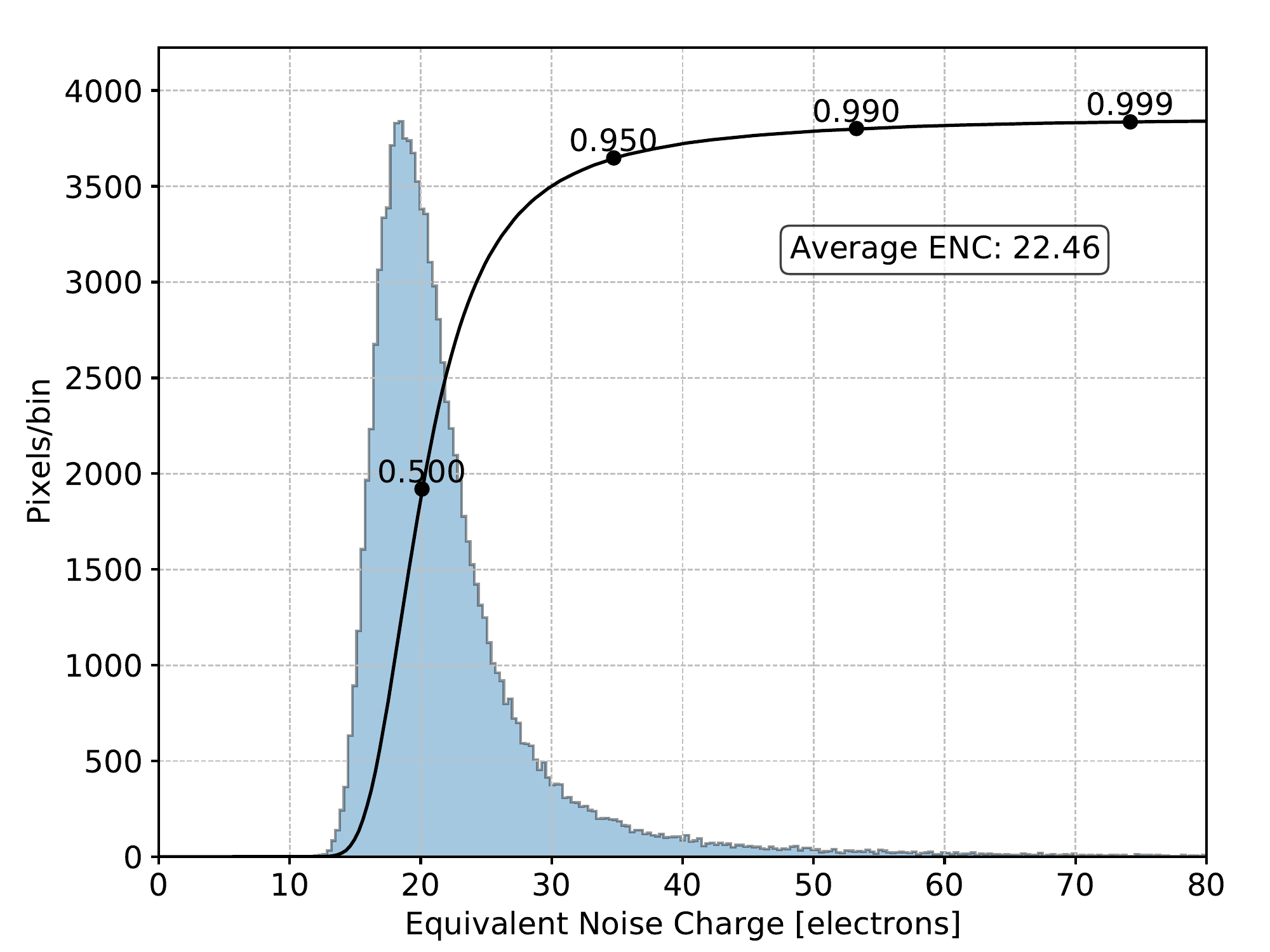}
    \caption{Distribution of the equivalent noise charge across the pixels for one of the 
    flight detectors. The annotated black line represents the cumulative distribution of the 
    pixel noise, with a few representative quantiles indicated by the black dots.}
    \label{fig:enc_distribution}
\end{figure}


We measure the noise of each pixel using the ASIC internal charge injection system and looking at the non-triggering pixels in the ROI.
Figure~\ref{fig:enc_distribution} shows a typical distribution of the equivalent 
noise charge. The average value is $\sim 22.5$~electrons rms, with only about one pixel in a
thousand exceeding $75$~electrons, and no significant spatial pattern across the active surface.
Due to the double-readout strategy used for nominal science acquisitions, the effective
noise is a factor $\sqrt{2}$ higher; yet, even at a gas gain as low as $200$, we are
sensitive to the single primary electron at more than $6\sigma$.

\begin{figure}[htb!]
    \centering
    \includegraphics[width=\linewidth]{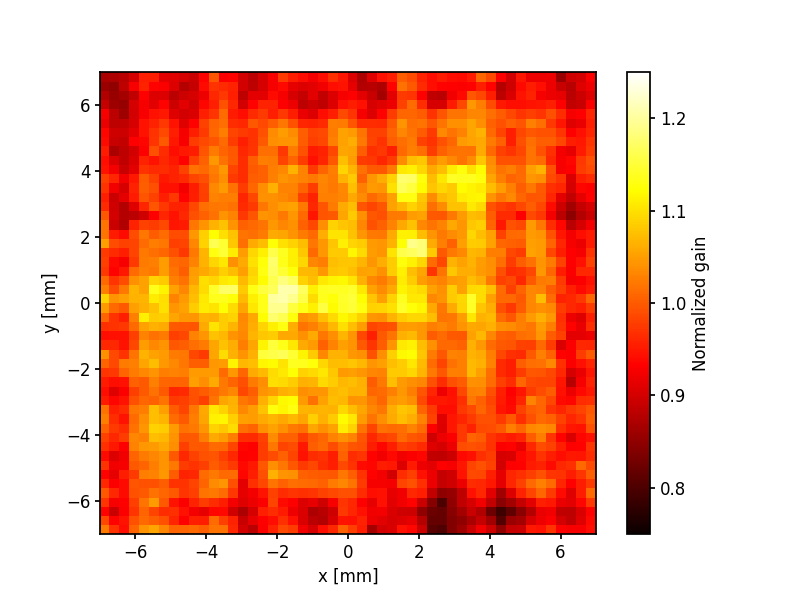}
    \caption{Map (in detector coordinates) of the normalized gain obtained 
    with a flat field with 5.9~keV X-rays from a $^{55}$Fe source. 
    (The average of the map is conventionally set to 1.) Most of the large scale
    disuniformities are due to small variations in the GEM thickness, while the 
    vertical and horizontal features are the imprinting of the manufacturing process
    described in Section~\ref{sec:gem_manufacturing}.}
    \label{fig:gain_map}
\end{figure}

The uniformity of response is typically measured with 
a flat field (i.e. a beam with almost uniform illumination covering the whole detector surface) using 5.9~keV
X-rays from a $^{55}$Fe source. The corresponding map of normalized gain---an 
example of which is show in Figure~\ref{fig:gain_map}---is also useful to verify the 
absence of defects (e.g., hot or dead spots) on the active surface.
The large-scale gain non-uniformity, which we ascribe to small variations in the thickness
of the GEM foil, is characterized by the dispersion of the gain values, and is typically
better than $10\%$ rms.
(This metrics for all the flight models is represented by the gray histogram in
Figure~\ref{fig:gpd_fe55_summary}.)
The vertical and horizontal patterns that are visible with a pitch of $1.8$~mm, on the
other hand, are a clear imprinting of the manufacturing process described in
Section~\ref{sec:gem_manufacturing} and are likely connected with a spatial modulation
of the microscopic hole inner structure, of which we have circumstantial evidence from
the optical metrology.

We further emphasize that, somewhat counter-intuitively, the reconstructed event direction
is extremely robust against these mid- and large-scale gain variations, 
although important to recover the ultimate energy resolution,
has little or no effect on the polarimetric response of the detector.

\subsection{Trigger Efficiency}

\begin{figure}[htbp!]
    \centering
    \includegraphics[width=\linewidth]{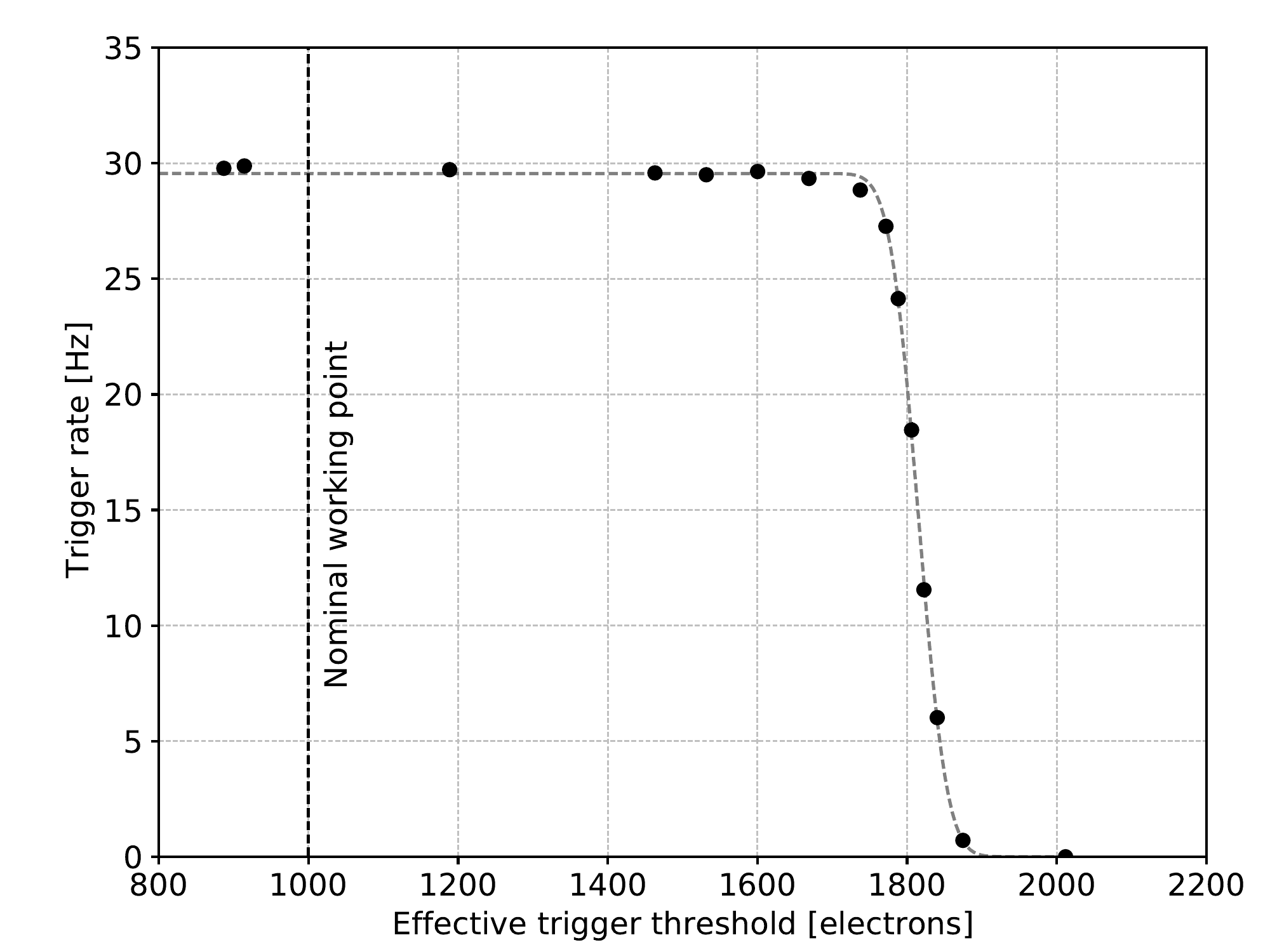}
    \caption{Event rate as a function of the trigger threshold with a $^{55}$Fe source
    irradiating the entire detector. The hardware trigger threshold is expressed in terms 
    of equivalent electrons, so that it can be compared directly with the average system noise.
    Note that the abrupt cutoff at $\sim 1800$ electrons is due to the limited dynamic
    of the trigger amplifier and does not correspond to the average energy released in a
    mini-cluster at the Bragg peak of the track (which is in excess of $5000$ in these
    units).}
    \label{fig:trg_efficiency}
\end{figure}

We typically operate the GPD at an effective trigger threshold of $\sim 1000$~electrons
of signal counts, which allows us to achieve a noise trigger rate $\ll 1$~Hz with only
handful of noisy pixels masked---typically less than one in 10,000, or the very far outliers in 
the distributions shown in Figure~\ref{fig:enc_distribution}. This ensures full 
efficiency with a fairly large margin, as illustrated by the threshold scan
in~Figure~\ref{fig:trg_efficiency}.

Since the trigger mini-clusters involved in the trigger are primarily those produced
by the Bragg peak of the track, and the latter is energy-independent, the metrics shown 
in Figure~\ref{fig:trg_efficiency} apply equally well to the entire IXPE energy
band. As mentioned in Section~\ref{sec:asic}, the padding in the definition of the
region of interest ensures that the beginning of the track, with a comparatively lower
ionization density, is captured in the readout even if the corresponding pixels do not
trigger up to the highest energies of interest.

\subsection{Pulse Height Analysis and Energy Resolution}

\begin{figure}[!bht]
    \centering
    \includegraphics[width=\linewidth]{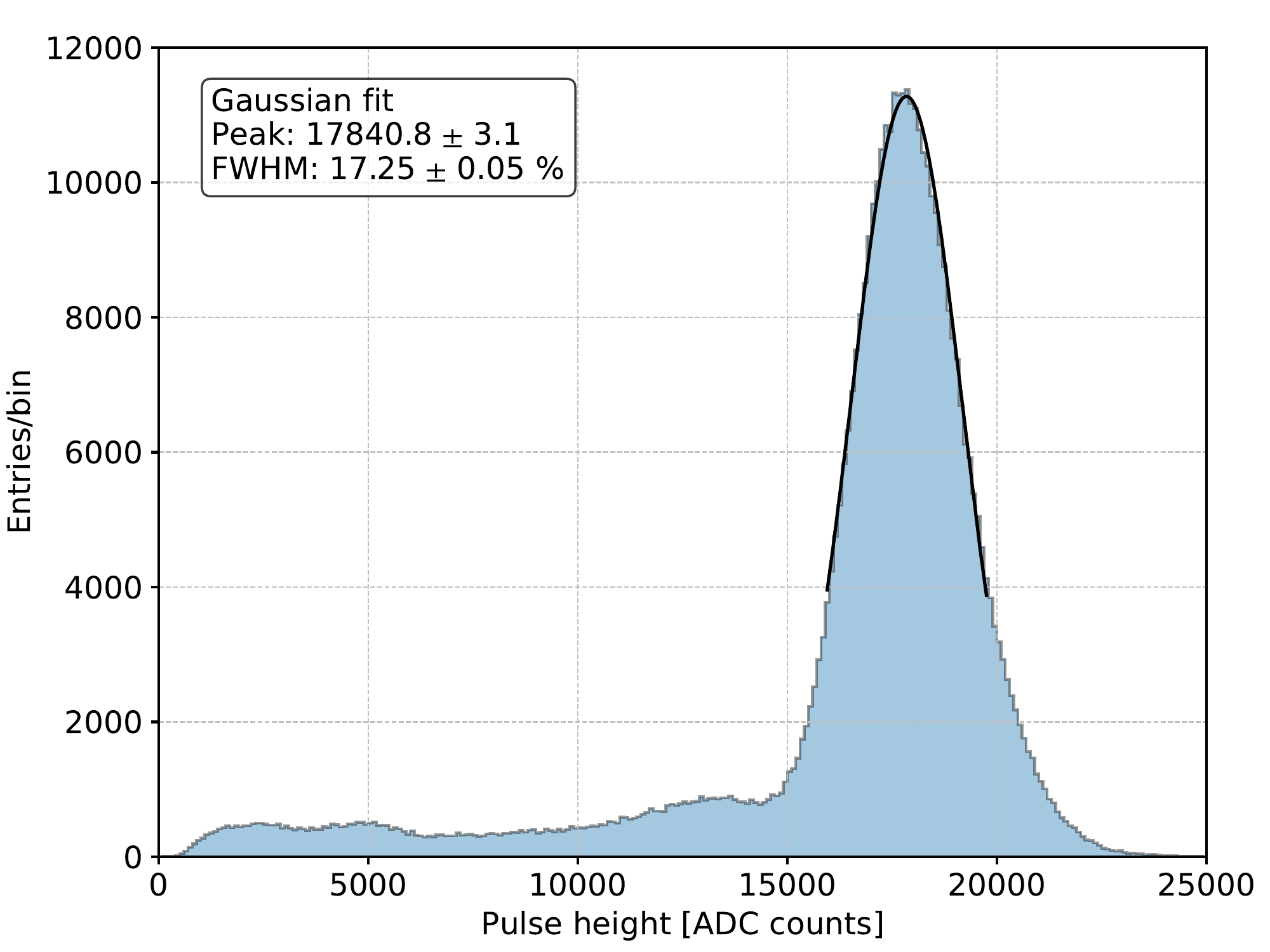}
    \caption{Typical GPD pulse-height spectrum, corrected for the non-uniformity of response,
    from a flat field at 5.9~keV. The energy resolution is $\sim 17$\% FWHM.
    The left tail of the distribution is due to events converting in the window or
    in the upper copper layer of the GEM, and is relatively less prominent at lower
    energies.}
    \label{fig:pha_spectrum}
\end{figure}

\begin{figure}[bt!]
    \centering
    \includegraphics[width=\linewidth]{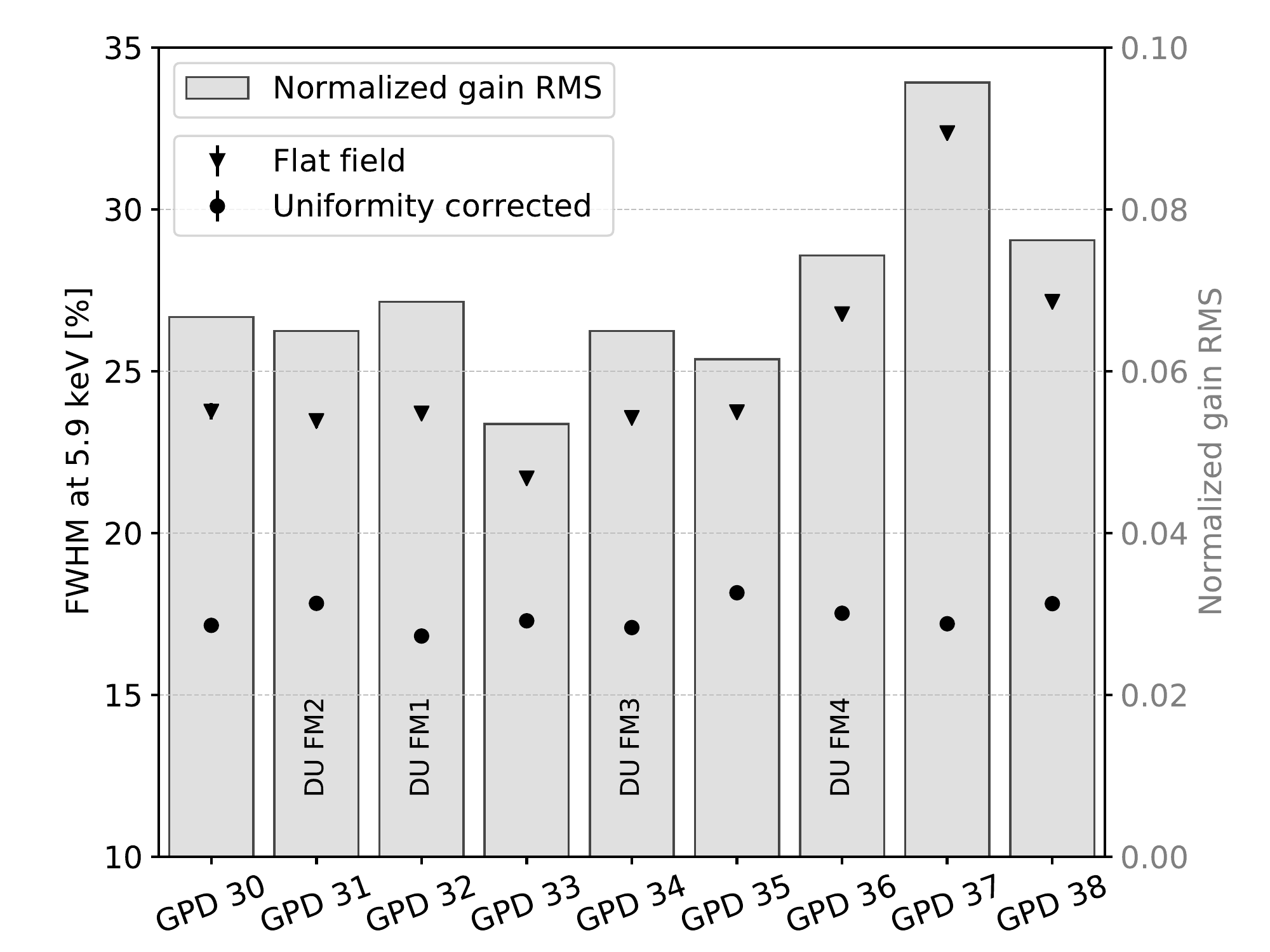}
    \caption{Summary plot of the GPD energy resolution at 5.9~keV. 
    The histogram shows the gain map dispersion, while the points are the energy resolution
    with and without the correction for gain non-uniformity.}
    \label{fig:gpd_fe55_summary}
\end{figure}

Figure~\ref{fig:pha_spectrum} shows a typical pulse height spectrum of a $^{55}$Fe source,
corrected for the spatial non uniformity of response. The energy resolution is of the order
of $\sim 17$\% FWHM at 5.9~keV for all the flight detectors, as shown in 
figure~\ref{fig:gpd_fe55_summary}, 
scaling approximately with the square root of the energy.
We have monitored the energy resolution for all the flight detectors over temporal baselines
of years without detecting significant variations, which is indirect evidence of the
leak-tightness of the detectors and of the quality of the assembly---as any small electro-negative
contaminants in the gas cell would have catastrophic consequences on the energy resolution.

Not all the events triggering the readout originate in the active gas
volume---occasionally photons absorbed in either the entrance window (and particularly in the
50~nm inner aluminum layer) or on the upper copper face of the GEM can produce a photoelectron that leaves the passive
material depositing enough energy in the gas cell. Such events, characterized by incomplete charge
collection, account for the vast majority of the low-energy tail in the pulse-height spectra.
The relative frequency of these passive conversion (Figure~\ref{fig:pha_spectrum}), estimated
from a detailed Monte Carlo simulation of the detector, is less than 2\% at 2~keV, increasing with
energy to almost 15\% at 8~keV, and needs to be modeled and calibrated using ground data 
in order to have a detailed description of the instrument response functions---the effective
area, the energy dispersion and the polarimetric response.

\subsection{Polarimetric Response}

\begin{figure}[!htb]
    \centering 
    \includegraphics[width=\linewidth]{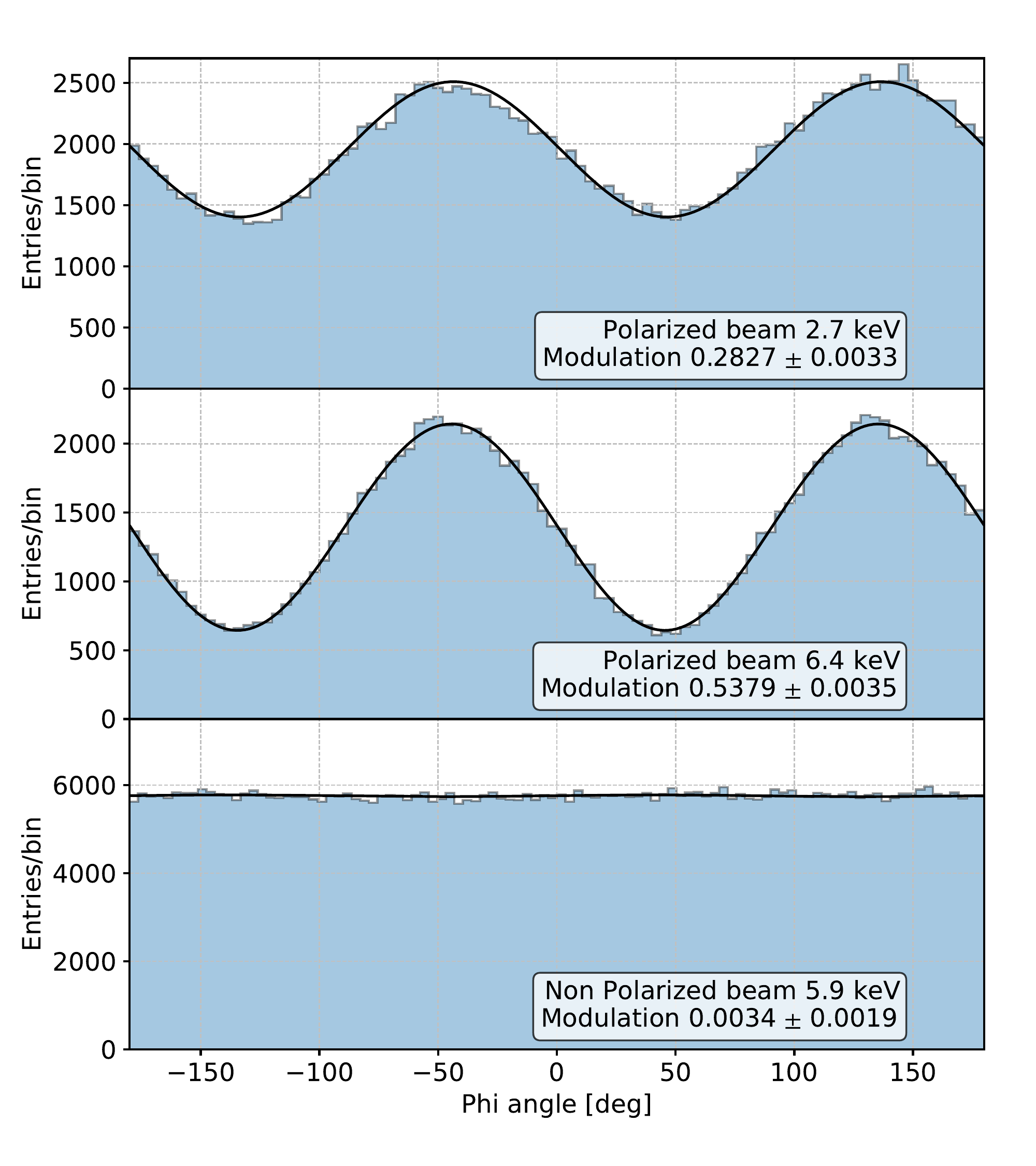}
    \caption{Example of modulation curves for 2.7 keV and 6.4 keV, $\sim 100\%$ polarized
    and 5.9~keV un-polarized X-rays. We note that the underlying selection cuts used
    are not necessarily identical to those that will be used for the analysis of flight data,
    and, as a consequence, the final modulation figures might be slightly different.}
    \label{fig:modulation_curve}
\end{figure}

\begin{figure}[!hb]
    \centering
    \includegraphics[width=\linewidth]{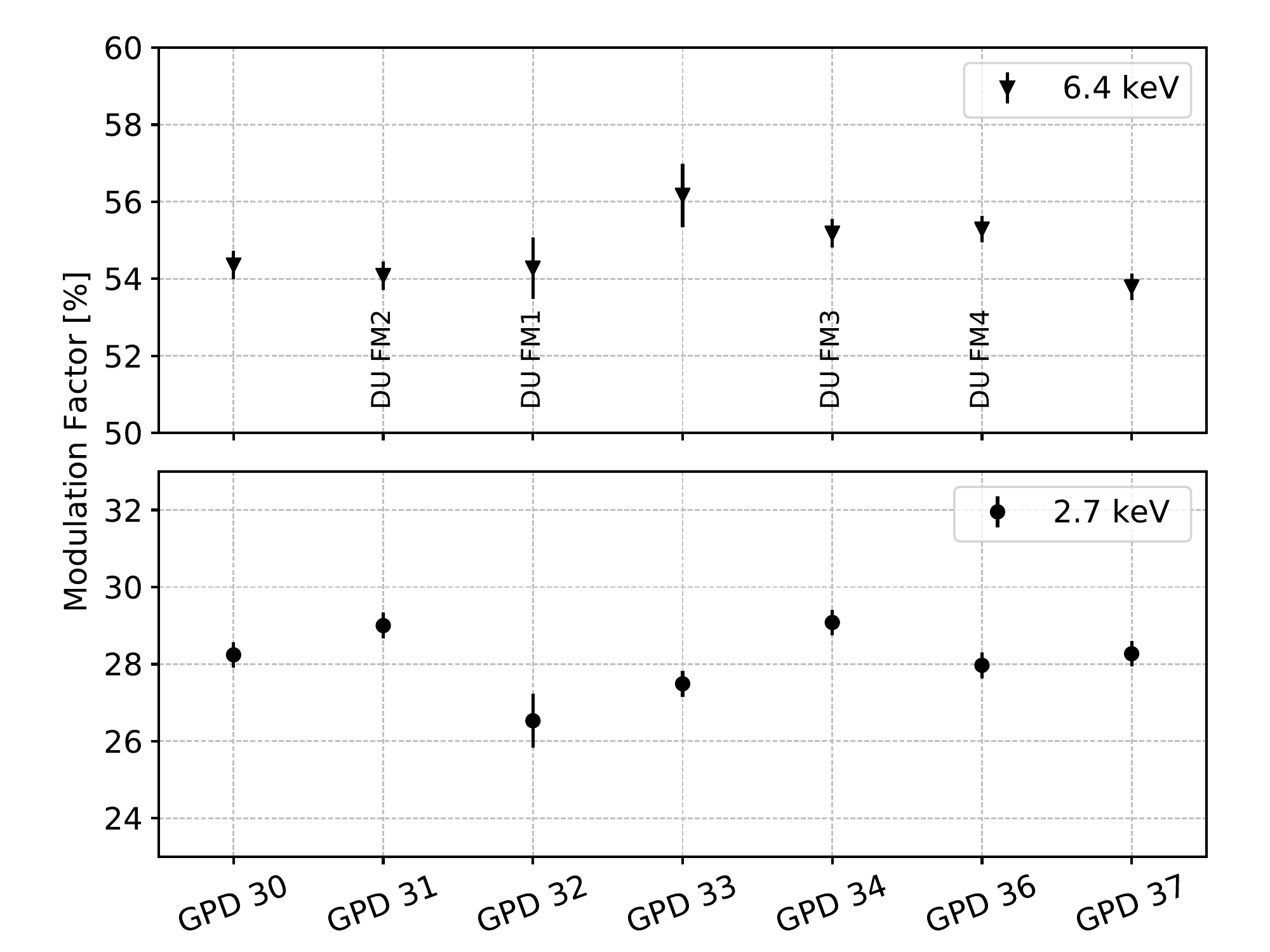}
    \caption{Modulation factor at 2.7 keV and 6.4 keV for all the GPDs tested during acceptance
    campaign of the flight models. (Note that GPDs~35 and 38 are not included, as they
    did not undergo a full acceptance test, and therefore were not included in the pool
    of candidates for the flight detector units, due to schedule constraints.)
    Differences in the test setup have been properly accounted for in order to allow for
    a fair comparison between different units.}
    \label{fig:modulation_factor}
\end{figure}

The modulation factor is routinely measured at two discrete energies (2.7 and 6.4~keV),
and for a set of different points on the active surface, for all the flight detectors
as part of the standard acceptance tests (prior to the integration of the
detector units). Figure~\ref{fig:modulation_curve} shows the typical modulation curves,
for 2.7 and 6.4 keV, $\sim 100\%$ polarized and 5.9~keV unpolarized X-rays.
The polarimetric response is fairly uniform across different detectors, as show in 
Figure~\ref{fig:modulation_factor}. The actual calibration of the flight detector 
units is significantly more extensive, and will be covered in details in a separate 
paper.

While the GPD polarimetric response to un-polarized radiation shows little or no sign
of systematic effects at 5.9~keV (i.e., the bottom modulation curve in
Figure~\ref{fig:modulation_curve} is statistically consistent with a uniform distribution),
we found a small but significant spurious modulation at lower energies, varying across the active 
detector surface, that needs to be properly calibrated and corrected to meet the IXPE 
science goals. We shall discuss the instrumental origin of such spurious modulation, that we
circumstantially traced back to the microscopic structure of the GEM holes, in 
Section~\ref{sec:systematics_spuriousmod}. The detailed calibration and correction strategy 
have been performed; however, a detailed description is beyond the scope of this paper and will be provided elsewhere. (See also Section~\ref{sec:systematics}.)

\subsection{Spatial resolution}

The GPD capability of measuring the X-ray impact point is strictly 
related to the photoelectron track imaging and polarimetric response. 
For this reason we did not measure the spatial resolution directly,
but relied on the result with polarized beam to indirectly verify
the adherence to the expected performance. 
This choice has been dictated by the tight schedule of the mission. 
Dedicated measurements are performed during the calibration campaign
(including tests with the optics) and will be discussed in separate papers. 
Here we can anticipate that the expected resolution of 
$\sim 100~\mu$m (half-power diameter, on-axis, 
almost flat in energy) is met, equivalent to 5 arcsec at 4 m,
leading to a telescope point spread function (PSF)
dominated by the optics' resolution.

\subsection{Dead Time}

The dead time per event $T_d$ depends on the specifics of the readout sequence and 
can in general be factored into two different terms---one constant and one proportional
to the number of pixels $n_\text{pix}$ in the region of interest:
\begin{align}\label{eq:deadtime_parametrization}
    T_d = d + m \; n_\text{pix}.
\end{align}
The two coefficients depend on the particular readout settings of the back-end electronics:
roughly speaking, $m$ is mainly determined by the clock period of the serial readout while
$d$ relates to the timing constraints imposed by the ASIC for a correct readout, and
particularly the fixed delay ($\geq 400~\mu$s) needed between two successive readouts.
(Both figures further depend on the number of additional event readouts used for the pedestal
subtraction).
In nominal data taking configuration $d \sim 750~\mu$s and $m \sim 600$~ns per pixel,
yielding an average dead time per event slightly in excess of 1~ms for a typical ROI of
500 pixels.

\begin{figure}
    \centering
    \includegraphics[width=\linewidth]{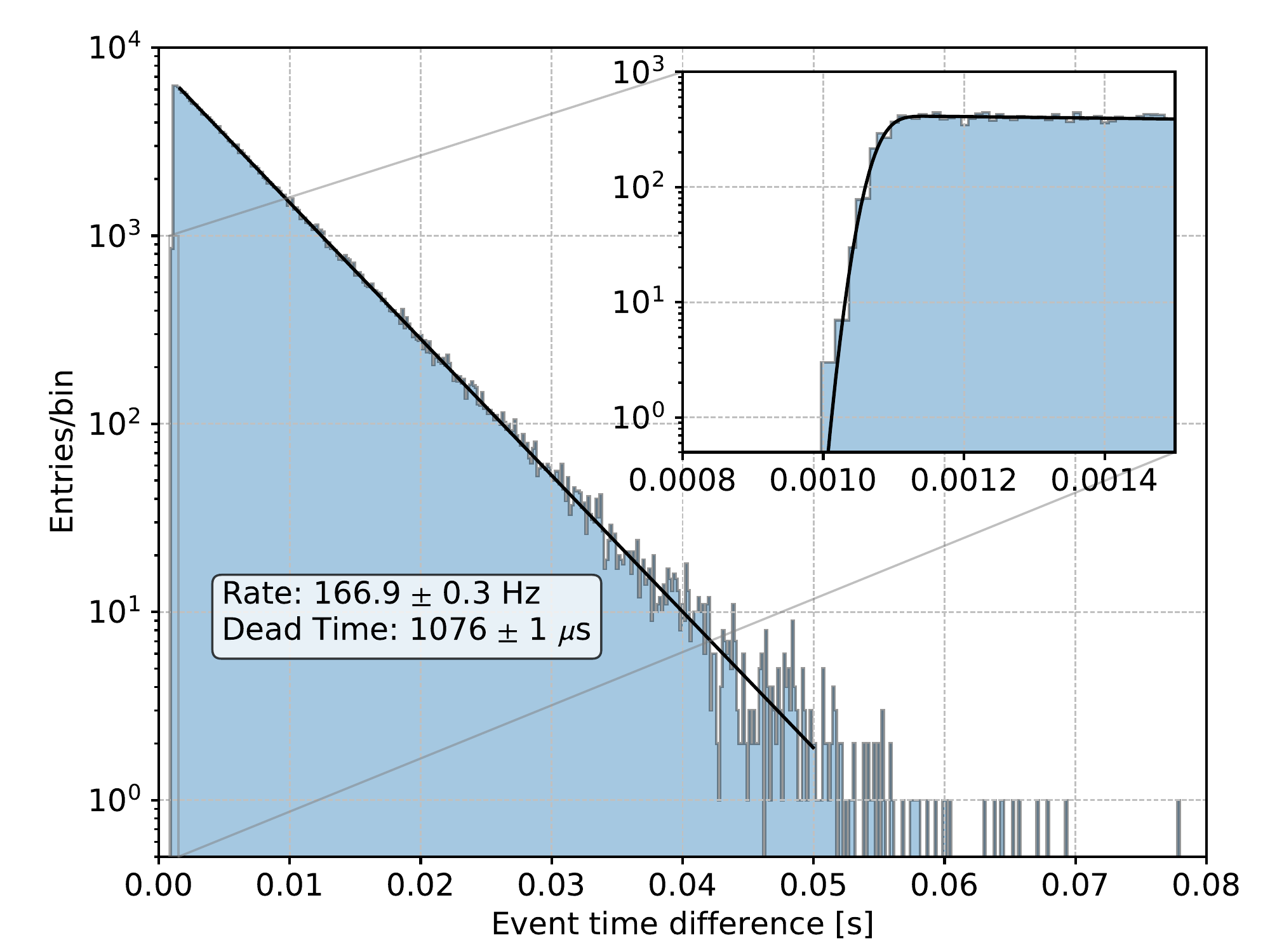}
    \caption{Distribution of the time differences between successive events, for a 
    Physics data acquisition at 2.7~keV at $\sim$170~Hz average rate in nominal configuration.
    The dead-time dispersion reflects the distribution of the ROI size at the beam energy.}
    \label{fig:time_difference}
\end{figure}

Figure~\ref{fig:time_difference} shows the time difference between two consecutive events
for monochromatic 2.7 keV X-rays (representative of the average energy for a typical
astronomical observation) in nominal data-taking configuration. The average dead-time per
event is in line with our parametrization~\Eqref{eq:deadtime_parametrization}. To put things in
context, this will enable observations of the Crab Nebula, which is the archetypal prototype of
our bright sources, yielding $\sim 80$~counts per second per detector unit, 
at $< 10\%$ overall dead-time.

\section{Systematic Effects}
\label{sec:systematics}

Long-term operations and in-depth performance characterization of the many GPDs developed for
the IXPE mission unveiled three different sources of systematic effects which add up to
the nominal detector operation as presented so far and were never fully documented in previous
publications.
We find that such effects have different nature and magnitude, and can be strongly constrained
by either dedicated calibrations, offline data analysis or specific operation modes of the
telescope, with a controlled, small effect on the IXPE scientific throughput.

\subsection{Low-energy Azimuthal Response}
\label{sec:systematics_spuriousmod}

Early tests of engineering and qualification GPD models with low-energy, un-polarized pencil
beams indicated the presence of a residual modulation amounting to an average
amplitude of several \% at 2.7 keV, rapidly decreasing with energy, and varying over spatial
scales smaller than the PSF of the IXPE optics on the detector active surface
(i.e., at least down to a few hundred $\mu$m).
All the data we collected indicate that such modulation is stable in time, and not dependent
on the temperature, the GEM  gain, or the trigger rate, and can be therefore calibrated by
means of dedicated ground measurements, and subtracted 
from science observations at a level consistent
with the IXPE design sensitivity.

This systematic, uneven azimuthal response of the GPD, often referred to as
\emph{spurious modulation}, is hard to model from first principles as it results from a sum of
different sources whose magnitude is strongly related to the track shape.

A first type of effects show no measurable dependence on the position across the active area
of the detector and is therefore straightforward to correct.
\begin{itemize}[leftmargin=10pt]
    \item We found that the ASIC digital activity within the readout generates a tiny shift of
    the baseline for a definite subset of the pixels, not fully canceled by the pedestal
    subtraction, and manifesting itself as a series of vertical structures affecting
    the region of interest with an even-odd pattern of $\sim 10$ electrons amplitude. (This is to be 
    compared to an average pixel noise of $20$ electrons and an actual signal of thousands of
    electrons per pixel in the core of the track.)
    The ASIC layout incorporates some unavoidable design asymmetries (e.g., column-wise bias voltage distribution, horizontal readout, pixel and mini-cluster footprints) that could in principle
    account for such behavior. However, the effect is so small that it is necessary to stack thousands
    of track images to be able to even measure it.
    Yet, by virtue of its \emph{coherent} nature, this effect can produce a clear effect at the same frequency of the genuine signal at the level of a few~\% for low-ellipticity tracks, typical of
    low energy photons. We therefore measure it and entirely correct it in the offline software.
    \item When the serial clock approaches the characteristic settling time of the readout chip
    analog buffer, the shape of the modulation curve can be affected via effective cross
    talk between adjacent pixels. In such conditions, the signal at the output of the serial
    buffer is latched too early and this causes subtle deformations of the track image. 
    This effect is mitigated by limiting the readout frequency to about $\sim 5$~MHz
\end{itemize}

\begin{figure}
    \centering
    \includegraphics[width=\linewidth]{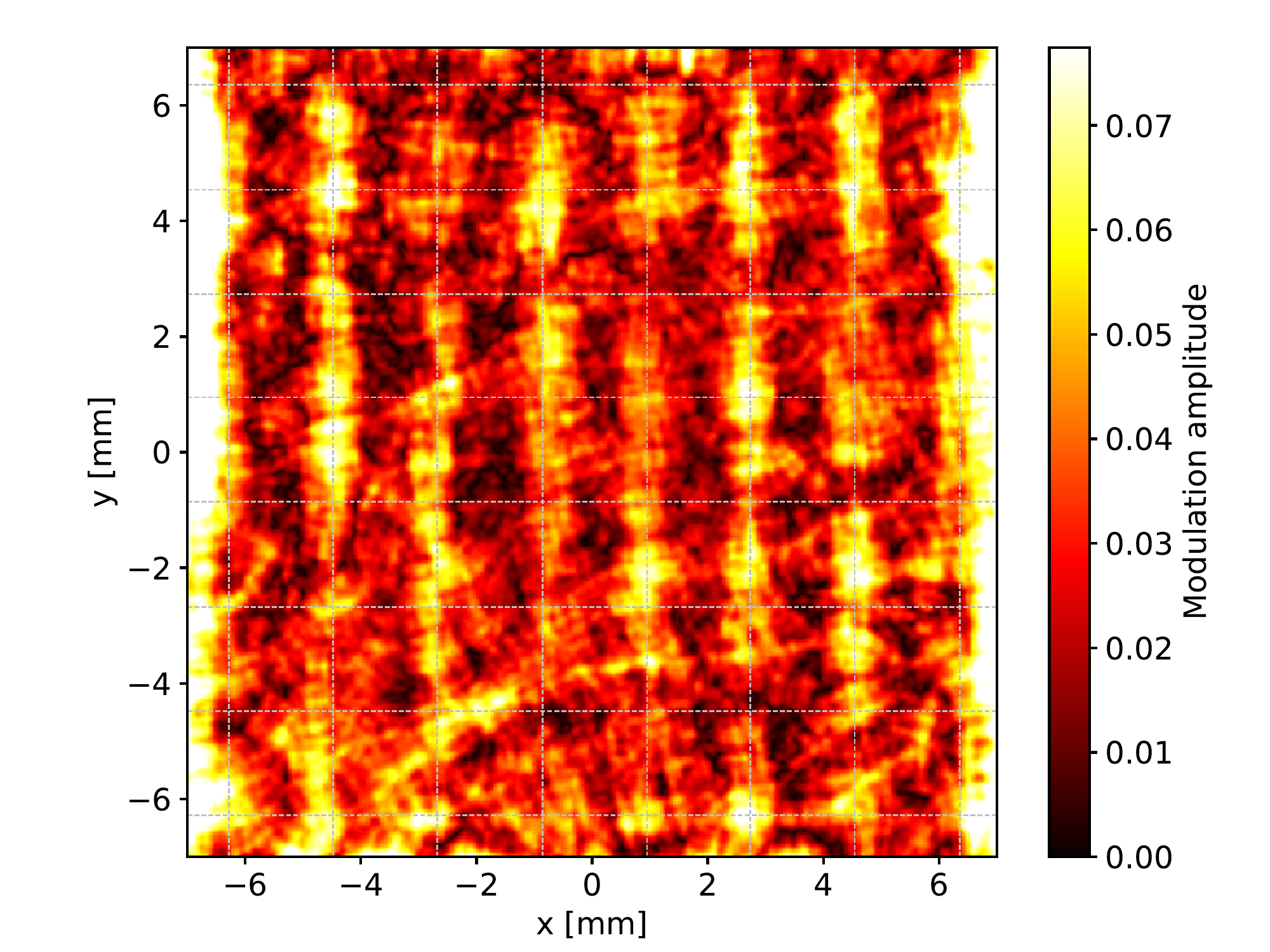}
    \caption{Modulation amplitude map obtained from an un-polarized X-ray source at 2.7~keV.
    The map was generated combining about 250~hours worth of data, for a total of $\sim 100$
    millions tracks, and smoothed with a 0.5~mm radius circular kernel.
    The dashed grid overlaid to the color plot corresponds to the known coordinates of the 
    laser sweep overlaps at the GEM manufacturing stage, see Section~\ref{sec:gem}, and 
    Figures~\ref{fig:gem}.
    (In order to emphasize the visual prominence of the structures, the ASIC effects
    have not been corrected.)}
    \label{fig:spurious_modulation_map}
\end{figure}

A second type of asymmetric GPD response shows a clear position dependence in high statistics
maps recorded from un-polarized beams, such as the one shown in Figure~\ref{fig:spurious_modulation_map}.
By analyzing these images, we firmly track back the source of this behavior to the GEM manufacturing.
\begin{itemize}[leftmargin=10pt]
    \item The features in these modulation maps  precisely track the known coordinates of the
    vertical lines where adjacent laser-drilling sweeps overlap---in other words, the regular
    pattern in the map, with a 1.8~mm horizontal spacing, is a clear imprinting from the GEM
    manufacturing process.
    \item Similar maps from qualification model GPDs assembled with chemically-etched GEM foils
    show significant differences: (i) the modulation amplitude is considerably smaller when
    averaged over relatively large regions, although peak magnitudes can be similar; 
    (ii) no coherent pattern between different detectors was seen, clearly reflecting the
    different mechanisms used to drill the holes in the dielectric (laser-based, inherently
    regular, vs. chemical).
\end{itemize}

All the data we have collected unambiguously exclude gain non-uniformity in the GEM
as the cause of the spurious modulation. Although toy Monte Carlo simulations show that a stretch
of the \emph{effective} pitch of the GEM holes in one of the two orthogonal directions as
small as $\sim 1\%$ could explain the observed modulation, provided that it is characterized
by a coherence length of the order of the track size (a few hundred $\mu$m), the effect is
too small to be measured directly with optical scans.

Correcting similar effects by mapping the response of each GPD down to the spatial 
scales of the detector PSF is in principle possible, but prohibitively time-consuming when considering
the necessary statistical accuracy, the total number of flight units and the multiple energy
layers to be calibrated. (For reference, Figure~\ref{fig:spurious_modulation_map} alone required
250 hours of un-interrupted dedicated calibrations.)
The decision was then taken to smooth these effects by dithering the Observatory along the
line of sight for nominal science observations. The dithering effectively broadens the image of
the source in detector coordinates (in our case over several mm$^2$ for a point source) while
preserving the intrinsic angular resolution of the telescope. Averaging over a larger area is
beneficial as it decreases the average modulation amplitude thanks to partial cancellation
due to phase incoherence.
We shall address the details of the calibration procedure and its statistical treatment as
an effective background component for the observations of celestial sources, in an upcoming
companion paper.

\subsection{Rate-Dependent Gain Variations}

Although one of the primary reasons for the choice of the laser etching technology 
was the observation that standard-pitch ($> 100~\mu$m) GEM foils produced with this
process are largely immune to rate-dependent gain instabilities~\cite{TAMAGAWA2006418},
this desirable property does not carry over to the very fine pitch of the IXPE GEMs.
When the detector is irradiated, part of the charge from the avalanche can be temporarily
deposited onto the dielectric substrate of the GEM, modifying the configuration of the
electric field, and causing local (and reversible) changes in the gain. 
This was actively investigated through the development phase of the mission, and mitigated
by fine-tuning specific steps of the GEM manufacturing process. Nonetheless, a residual
effect at the $\sim 10\%$ level is still present in the IXPE flight units.

The basic phenomenology of this \emph{charging} is a definite decrease of the GEM gain, when
the detector is irradiated at a sufficiently high rate. The asymptotic gain value and the 
time-scale associated to the gain variations both depend on the input energy flux.
As the charge trapping is not permanent, a competing \emph{discharging} process, with a much
longer associated time scale, is continuously at play, causing the gain to recover,
provided that the input energy flux is low enough.
We have developed a simple phenomenological model of the effect, described in~\ref{sec:charging},
that allows us to predict and correct the gain variations, based on the energy flux measured
by the GPD itself as a function of time and position.

\begin{figure}[b!]
    \centering
    \includegraphics[width=\linewidth]{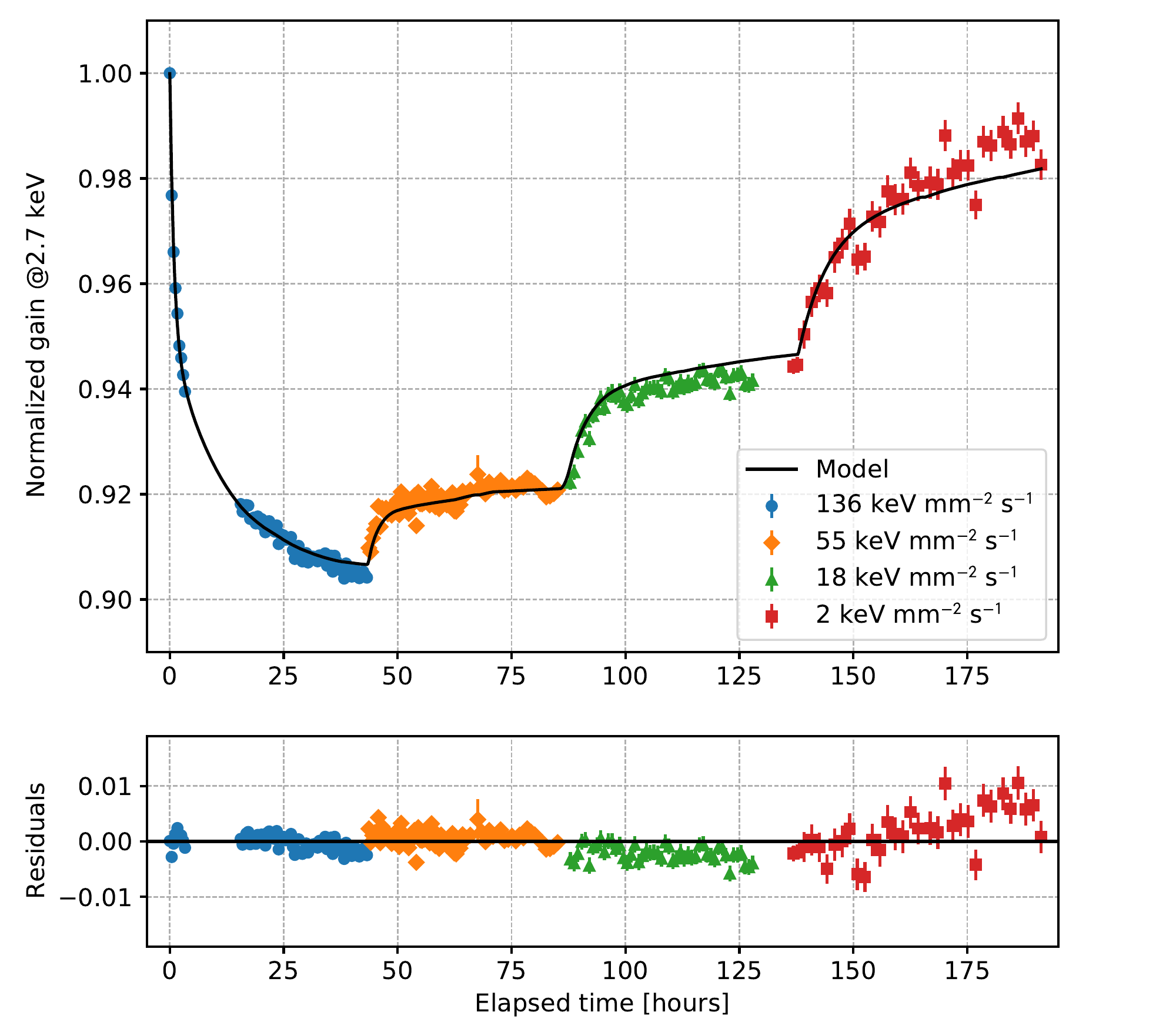}
    \caption{Modeling of the charging effects. The data points represent the relative 
    gain as a function of time, while the detector is being irradiated with a 
    monochromatic X-ray beam at different rates (different colors indicate different
    values of the corresponding energy flux). Note that the asymptotic gain and 
    the typical time scale for the variation is different for each setting. The
    solid line is the prediction of the model described in~\ref{sec:charging}, where the
    energy flux as a function of time measured by the GPD has been integrated
    self-consistently. (We emphasize this is not a piece-wise fit to the data points.)
    The residuals are between $\pm 1\%$ over 200 hours and two orders of magnitude of
    variation of the input energy flux.}
    \label{fig:data_vs_model_du2}
\end{figure}

We verified the accuracy of this description by irradiating the detector with X-ray sources
at different energies and rate per unit area. Figure~\ref{fig:data_vs_model_du2}
shows an example of a data acquisition performed for the purpose of studying the
charging and extracting the model parameters necessary for its dynamic correction.
Although typical X-ray celestial sources can induce about 
a few~$\%$ charging effect, 
as illustrated in \ref{sec:charging}, IXPE will also observe brighter sources throughout its
operations. Consequently, gain variations over very different time scales, spanning from minutes
to days, are to be expected. We shall rely on a dynamic correction of such variations based
on our phenomenological model, tuned for each detector unit using dedicated ground
calibrations, and informed by periodic on-orbit calibrations.

\subsection{Secular Pressure Variations}

One unexpected realization of the development of the IXPE instrument was the fact that
the pressure in the GPD gas cell decreases with time, over time scales of months,
with an overall asymptotic deficit of about $\sim 150$~mbar, compared to the nominal
800~mbar at filling time.

\begin{figure}[htb!]
    \centering
    \includegraphics[width=\linewidth]{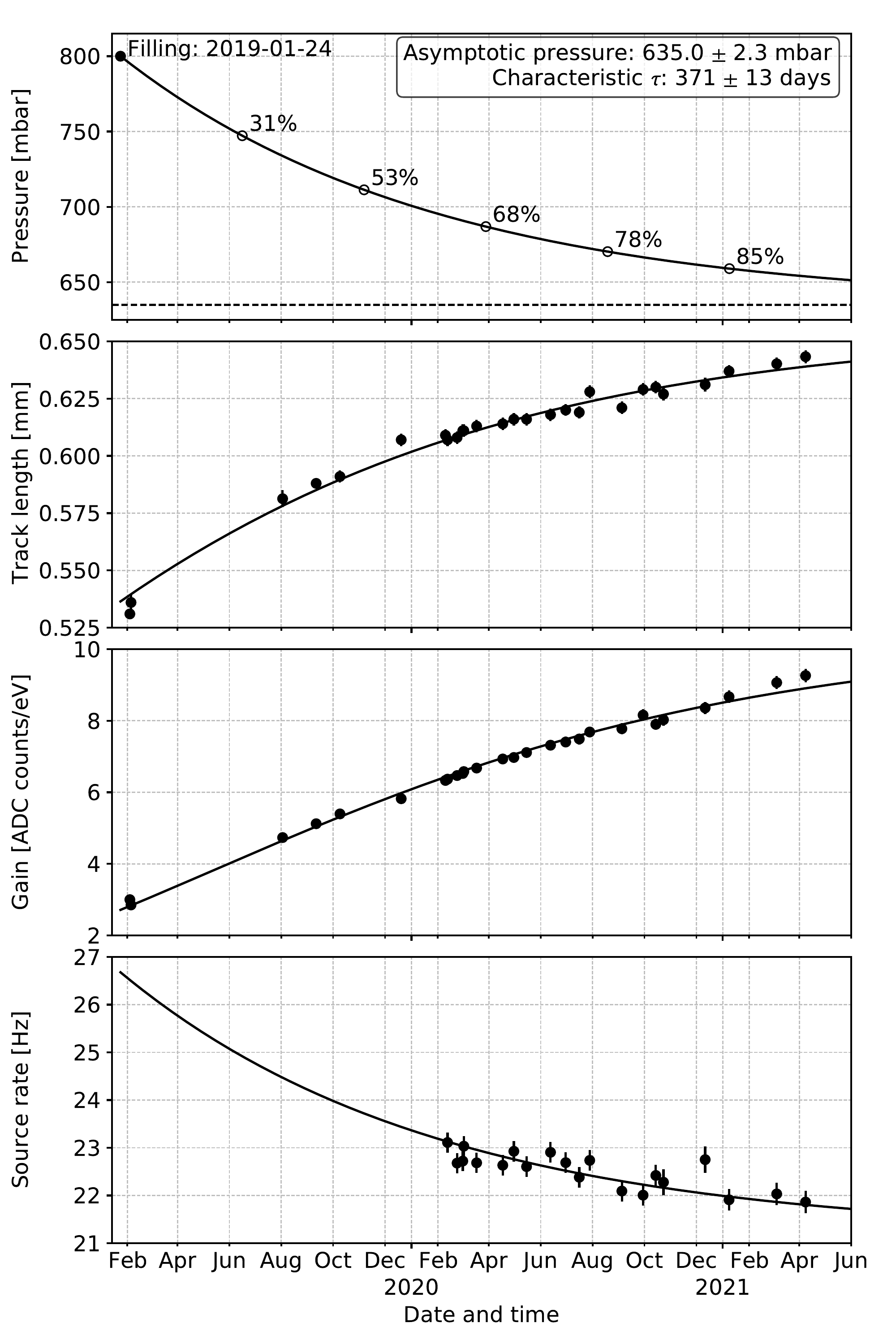}
    \caption{Secular pressure variation summary for GPD 30. The solid line on the top
    panel is the best-fit pressure model based on the change with time of the three
    proxy that we monitor: the gain, the track length, and the trigger rate from a 
    reference source. (The latter has been corrected for the radioactive decay.)}
    \label{fig:gpd_secular_summary}
\end{figure}

The first indirect hint of this phenomenon was a slow change with time of the gas gain,
that was then connected to a deficit of internal pressure based on the calibration of the
absolute quantum efficiency, which will be described in a subsequent paper.
While we have no direct measurement of the internal pressure of the gas cell, our continuous 
monitoring of key detector parameters for the flight detectors consistently support this
scenario: the average track length (and the modulation factor as a consequence) increase with
time, while the event rate from a reference source decreases accordingly, as shown in
Figure~\ref{fig:gpd_secular_summary}.
Furthermore, metrological measurements independently confirm that the vertical position of
the center of the beryllium window changes continuously due to the pressure reduction
(this is further discussed in~\ref{sec:secular_evolution}).
The effect tends to saturate with time---and, in fact, all the detectors installed in the 
flight detector units will be within a few \% from the asymptotic pressure value by the 
time IXPE is launched.

Although the root cause of the effect is unknown, it is clear that the phenomenon is not due
to a real leak between the gas cell and the external world (which is by the way excluded by
the leak tests performed at various stages of the GPD assembly) because the energy resolution
of all the flight detectors, with no exception, does not show any sign of worsening with time.
Additionally, we have experimental indications that the effect is largely reversible by
heating the gas cell at high ($\sim 100^\circ$~C) temperature---and therefore, in all
likeness physical, (as opposed to chemical) in nature.

As further discussed in~\ref{sec:secular_evolution}, we devised a simple analytical model
describing the pressure variation with time, that is adjusted to data on a detector-by-detector
basis by means of a simultaneous fit to the three pressure proxies that are regularly
monitored in time: the gain, the track length and the trigger rate from a reference
radioactive source. Any residual evolution in flight will be properly corrected using this
very model, which is also used to scale all the relevant performance metrics from the 
values measured at the time of the ground calibrations.

We emphasize that the impact of this pressure variation on the polarimetric sensitivity
is limited. On one hand, the detector quantum efficiency scales linearly with the pressure,
causing a net loss of effective area. On the other hand, the modulation factor increases as
the pressure decreases, owing to fact that the tracks become more elongated. The net effect of
these two competing processes at play is that the relative loss of sensitivity, expressed as
the broadband minimum detectable polarization for a typical source spectrum, is less than
2\% when going from 800 to 650~mbar. This is qualitatively consistent with the sensitivity
scaling across the phase space shown in Figure~\ref{fig:gpd_design_tradeoff}.

\section{Conclusions}

Gas Pixel Detectors, proposed nearly twenty years ago as a  revolutionary tool to access X-ray polarization properties in the photo-electric domain, have now reached full maturity to enable first-time, high-sensitivity  observations of astrophysical sources at the $\%$ level for several source classes.

This paper presents the design choices, the main steps in the integration and test flow of the detecting
elements, and the resulting system performance in terms of polarization sensitivity, imaging and
spectroscopic capabilities in the context of the NASA IXPE mission.

For the first time, we also discuss the main sources of systematic uncertainties that emerged along
the qualification of this technology for space, and their implications on operations of the IXPE mission.

\section{Acknowledgements}
The Italian contribution to the IXPE mission is supported by
the Italian Space Agency (ASI) through the contract ASI-OHBI-2017-12-I.0,
the agreements ASI-INAF-2017-12-H0 and ASI-INFN-2017.13-H0, and its
Space Science Data Center (SSDC), and by the Istituto Nazionale di
Astrofisica (INAF) and the Istituto Nazionale di Fisica Nucleare (INFN) in Italy.
T.~Tamagawa aknowledges support from the JSPS KAKENHI Grant Number JP19H05609.

\appendix
\section{Modeling Rate-Dependent Gain Variations}
\label{sec:charging}

We present the analytical effective treatment that we developed to describe and
correct the charging effect. Our results are largely in agreement with
the microscopic Monte Carlo simulation of the very same phenomenon presented
in~\cite{Hauer_2020}.

The simplest possible model encapsulating the observed charging development in amplitude and time can 
be cast in terms of the time-dependent accumulated charge per unit area $q(t)$ as
\begin{align}\label{eq:charging_charge}
  \frac{dq(t)}{dt} =
  \overbrace{R(t) \alpha_c \left(1 - \frac{q(t)}{q_\text{max}} \right)}^\text{charge}
  - \overbrace{\frac{q(t)}{\tau_d}}^\text{discharge},
\end{align}
where $R(t)$ is the input charge flux per unit area, $\alpha_c$ is an adimensional
charging constant, and $\tau_d$ is a discharge time constant.

The one additional piece of information that we need to recast the model in terms
of measured quantities is the link between the accumulated charge and the gain variation.
Since the overall effect is $\sim 10\%$, which
corresponds to a variation of a fraction of a \% in terms of the effective electric field,
it is reasonable to assume a linear relation between the \emph{change in the relative gain} 
$g(t)$ and the relative accumulated charge:
\begin{align}
  g(t) = \frac{G(t)}{G_0} - 1 = - \delta_\text{max} \frac{q(t)}{q_\text{max}}.
\end{align}
With this new setting, equation~\ref{eq:charging_charge} can be rewritten as
\begin{align}
  \frac{d g(t)}{dt} =
  \overbrace{-\frac{F(t)}{k_c} \left[\delta_\text{max} + g(t) \right]}^\text{charge}
  + \overbrace{\frac{g(t)}{\tau_d}}^\text{discharge},
\end{align}
where $F(t)$ is the input energy flux per unit area of the source, measured in
units of keV~s$^{-1}$~mm$^{-2}$, $\delta_\text{max}$ is the maximum relative gain
excursion, reached when the accumulated charge saturates to its maximum value
$q_\text{max}$, and $k_c$ is a charging constant, measured in keV~mm$^{-2}$,
incorporating the $\alpha_c$ and $q_\text{max}$ parameters of our original model,
that we cannot measure separately.

Naive as it is, this very simple model can be solved analytically in the particular
case of constant input energy flux $F$, which allows us to derive a number of
interesting consequences. First of all, the quantity 
\begin{align}
    \tau_c(F) = \frac{k_c}{F}
\end{align}
acts as a charging time constant. Since $\tau_c$ is inversely proportional to $F$, 
the charging process is faster the larger the input energy flux.
The system evolves with an effective time constant
\begin{align}
    \tau_\text{eff}(F) = \left( \frac{1}{\tau_c}  + \frac{1}{\tau_d} \right)^{-1} = 
    \left( \frac{F}{k_c}  + \frac{1}{\tau_d} \right)^{-1}
\end{align}
toward an asymptotic relative gain variation given by
\begin{align}
    \delta(F) = -\frac{\delta_\text{max}}{\left( 1 + \frac{\tau_c}{\tau_d}\right)} =
    -\frac{\delta_\text{max}}{\left( 1 + \frac{k_c}{F\tau_d}\right)},
\end{align}
i.e., the overall asymptotic gain excursion increases with the input energy flux
and reaches $\delta_\text{max}$. Finally, there exists a \emph{critical} energy 
flux 
\begin{align}
    F_\text{critical} = \frac{k_c}{\tau_d}
\end{align}
separating the two fundamentally different regimes: when $F \gg F_\text{critical}$
the system is dominated by the charging processes and $\delta(F) \rightarrow{} \delta_\text{max}$
with a time constant $\tau_\text{eff} \rightarrow k_c/F$. On the other hand, when 
$F \ll F_\text{critical}$ the discharge process is dominating, and the  
gain tends to the unperturbed value ($\delta(F) \rightarrow 0$) with a time constant
$\tau_\text{eff} \rightarrow{} \tau_d$. Incidentally,
$\delta(F_\text{critical}) = -\delta_\text{max} / 2$.

\begin{table}[htbp!]
    \centering
    \begin{tabular}{p{0.4\linewidth}p{0.5\linewidth}}
    \hline
    Parameter & Typical Value\\
    \hline
    \hline
    $\delta_\text{max}$ & $\sim 0.1$\\
    $k_c$ & $\sim 10^5$~keV~mm$^{-2}$\\
    $\tau_d$ & $\sim 10^5$~s\\
    $F_\text{critical}$ & $\sim 1$~keV~mm$^{-2}$~s$^{-1}$\\
    \hline
    \end{tabular}
    \caption{Typical values of the charging parameters.}
    \label{tab:charging_paremeters}
\end{table}

When performing celestial observations, controlling the absolute energy scale 
is important for the measurement of the absolute source fluxes.
More importantly, an unbiased energy estimate on an event by-event basis is 
pivotal for the polarization measurement, 
as the modulation factor is energy dependent.

\begin{figure}[htbp!]
    \centering
    \includegraphics[width=\linewidth]{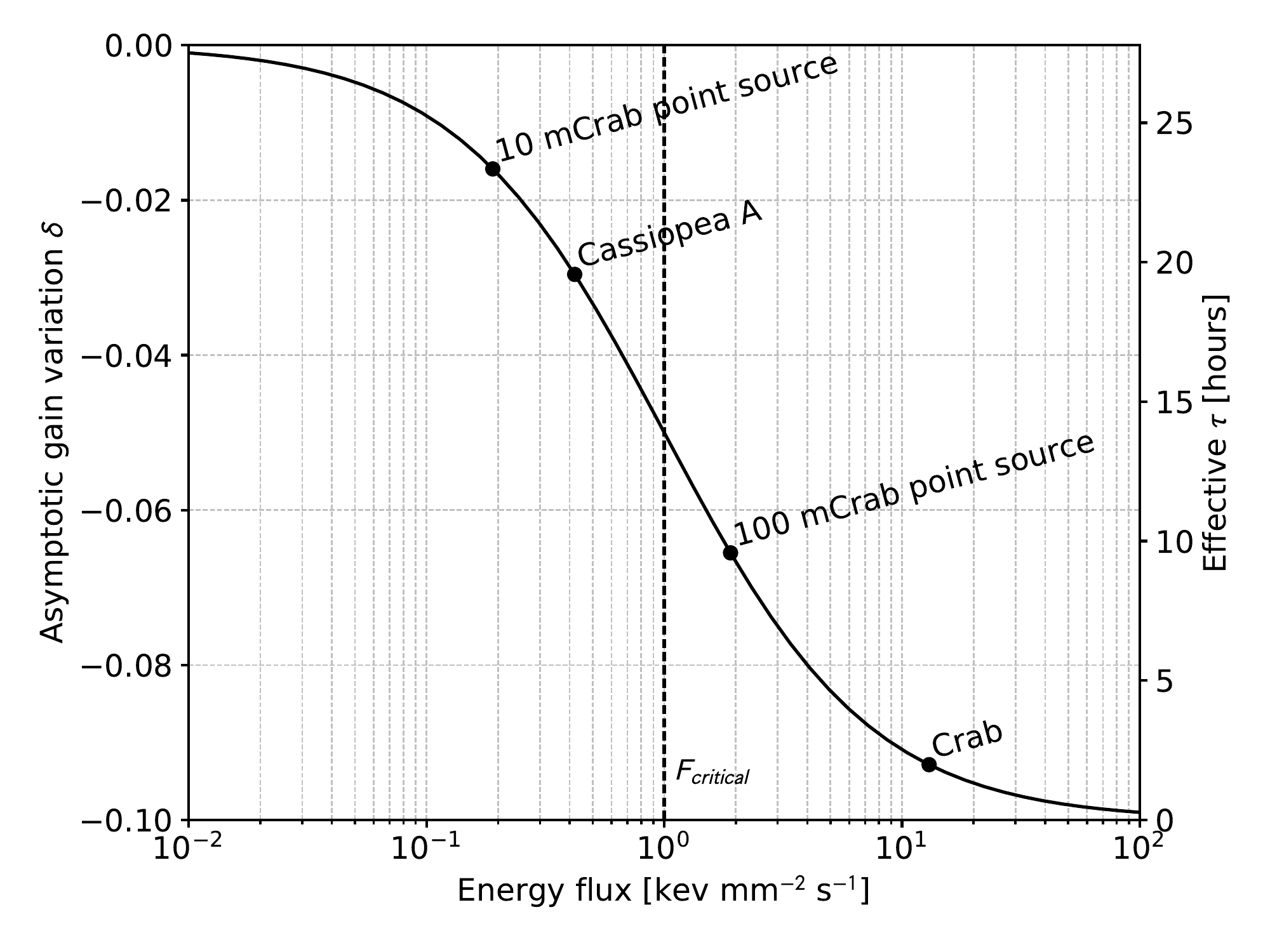}
    \caption{Asymptotic gain variation and associated time scale as a function of the input
    energy flux, for the representative set of parameters shown in Table~\ref{tab:charging_paremeters}.
    A sample of significant targets from the IXPE preliminary observing plan are overlaid
    for reference. The vertical dashed line represents the critical energy flux, setting the 
    natural scale for the magnitude of the charging effect on high-level science analysis.}
    \label{fig:charging_science_impact}
\end{figure}

Table~\ref{tab:charging_paremeters} shows the typical values for the charging parameters, 
while Figure~\ref{fig:charging_science_impact} shows the potential impact on different science observations.
For faint sources the gain variations will not be appreciable over the typical 
time scale for an observation, but for medium to bright sources the charging 
correction is definitely important.

\section{Modeling Pressure Variations}
\label{sec:secular_evolution}

In absence of direct measurements, we use mainly three proxies for inferring the
internal pressure of the gas cell: the (absolute or relative) quantum efficiency,
the average track length and the gas gain.

The quantum efficiency is, in principle, the most straightforward measurement of
the pressure. Measurements of the absolute quantum efficiency are somewhat complex
to perform, and are typically systematic-limited, which makes it impractical to use
them as a means for fine, long-term monitoring of the detector performance.
The quantum efficiency can be measured in relative terms, e.g., by measuring the event
rate with a reference radioactive source in a standard holder. This is a much simpler
measurement to do and has been indeed systematically exploited in order to investigate
the secular pressure variations.

Any change of the pressure in the gas cell will cause changes in the track topology,
e.g., through changes in the electron range and transverse diffusion. In principle
there are several (not independent) topological track quantities that can be used
for the purpose. Among them, the track length (defined via the moment analysis event reconstruction)
is the most robust against changes in the environmental conditions and the data acquisition
settings, and is empirically found to scale as
\begin{align}\label{eq:track_length_scaling}
    L(p) = L_0 \left( \frac{p}{p_0} \right)^{-\alpha_L}.
\end{align}
with $\alpha_L = 0.867$.

\begin{figure}[htbp!]
    \centering
    \includegraphics[width=\linewidth]{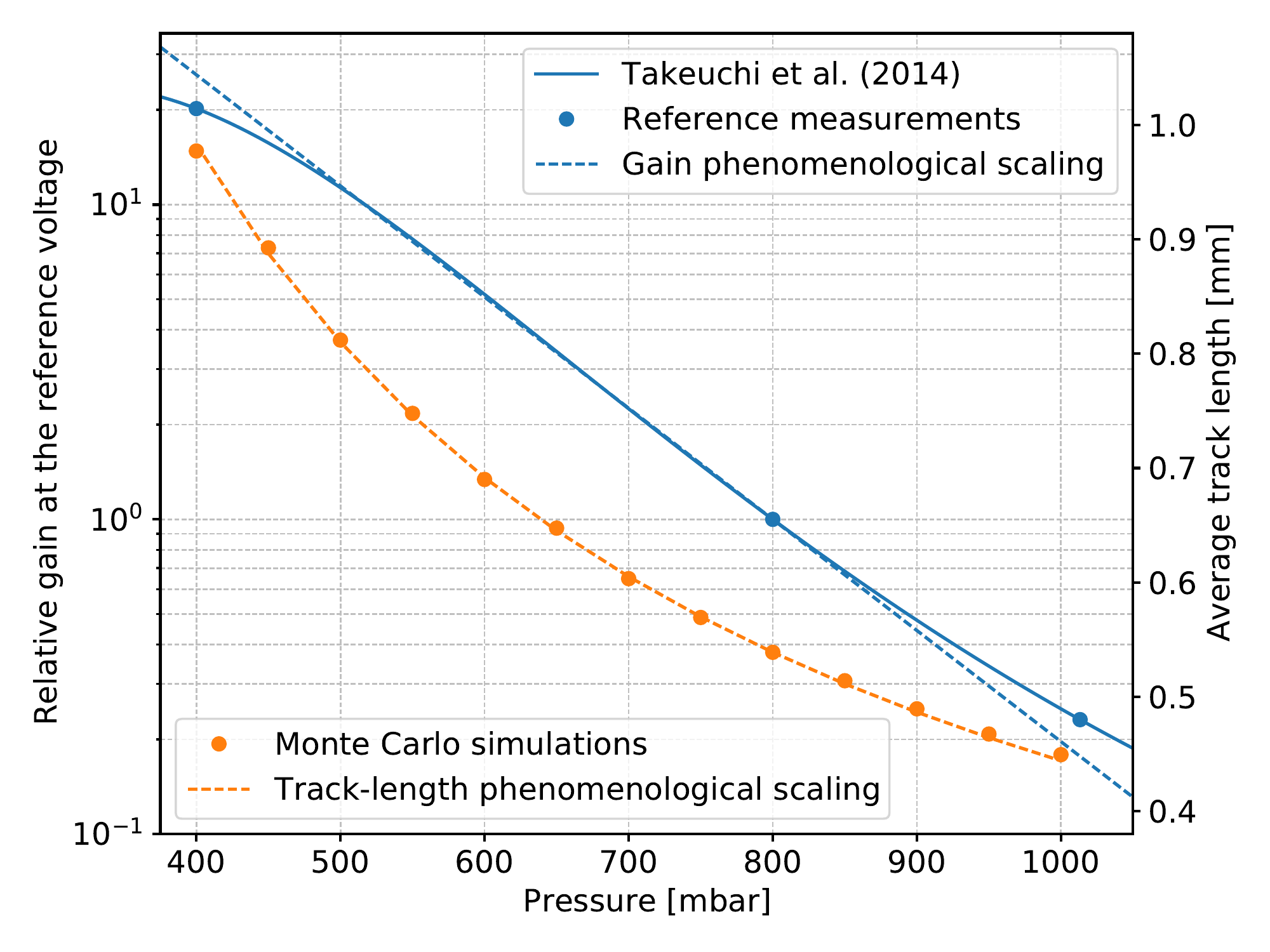}
    \caption{Scaling of the average track length at 5.9~keV and the relative gas gain used
    for the modeling of the secular pressure variations of the gas pixel detector.
    The points for the track length have been calculated through dedicated Monte Carlo 
    simulations and fitted to a power law~\Eqref{eq:track_length_scaling}.
    The basic scaling for the gain is taken from~\cite{Takeuchi_2014} and taylored to 
    our specific geometry by means of dedicated measurements at 400, 800 and $\sim 1000$ mbar.
    In the pressure range of interest (600--800~mbar) the phenomenological 
    parametrization~\Eqref{eq:gain_scaling} is in agreement with the underlying model to 
    about 1\%.}
    \label{fig:gpd_gain_pressure_scaling}
\end{figure}

The gas gain depends on the gas pressure at a fixed composition, and is a third,
independent proxy that we have customarily used to investigate the secular variations.
We emphasize that, compared to the former two quantities, the gain is somewhat more difficult
to use, as its dependence on the pressure is complex, and the pulse-height must be properly
re-scaled to a common high-voltage working point in order to compare different detectors.
In addition, the gain is known to show other kind of variations (e.g., due to the GEM charging,
or induced by changes in the environmental conditions) that need to be carefully controlled.
We use the parametrization in~\cite{Takeuchi_2014} for modeling the pressure dependence
of the gain, anchored to our detector geometry by means of dedicated measurements at three
different pressures, as shown in figure~\ref{fig:gpd_gain_pressure_scaling}.

From a purely phenomenological point of view, the scaling of the gain vs. pressure is
approximately exponential (at least locally), and the
\begin{align}\label{eq:gain_scaling}
    G(p) = G_0 \; \text{exp} \left\{-\frac{(p - p_0)}{p_\text{scale}} \right\},    
\end{align}
with $p_\text{scale} = 123$~mbar describes the full model to within 1\% around
$p_0 = 800$~mbar.

We model the time dependence of the pressure with a
two-parameter exponential function 
\begin{align}
    p(t; \tau, \Delta_p) = 
    p_0 - \Delta_p \left(1 - \text{exp}\left\{-\frac{(t - t_0)}{\tau}\right\} \right)
\end{align}
where $\tau$ and $\Delta_p$ are the time constant of the process and the asymptotic 
pressure loss, respectively. (In contrast, $p_0$ is assumed to be the nominal $800$~mbar
at the filling time $t_0$.) 

We recover $\tau$ and $\Delta_p$ from a combined fit of the three independent proxies,
as illustrated in figure~\ref{fig:gpd_secular_summary}.

\begin{figure}[htbp!]
    \centering
    \includegraphics[width=\linewidth]{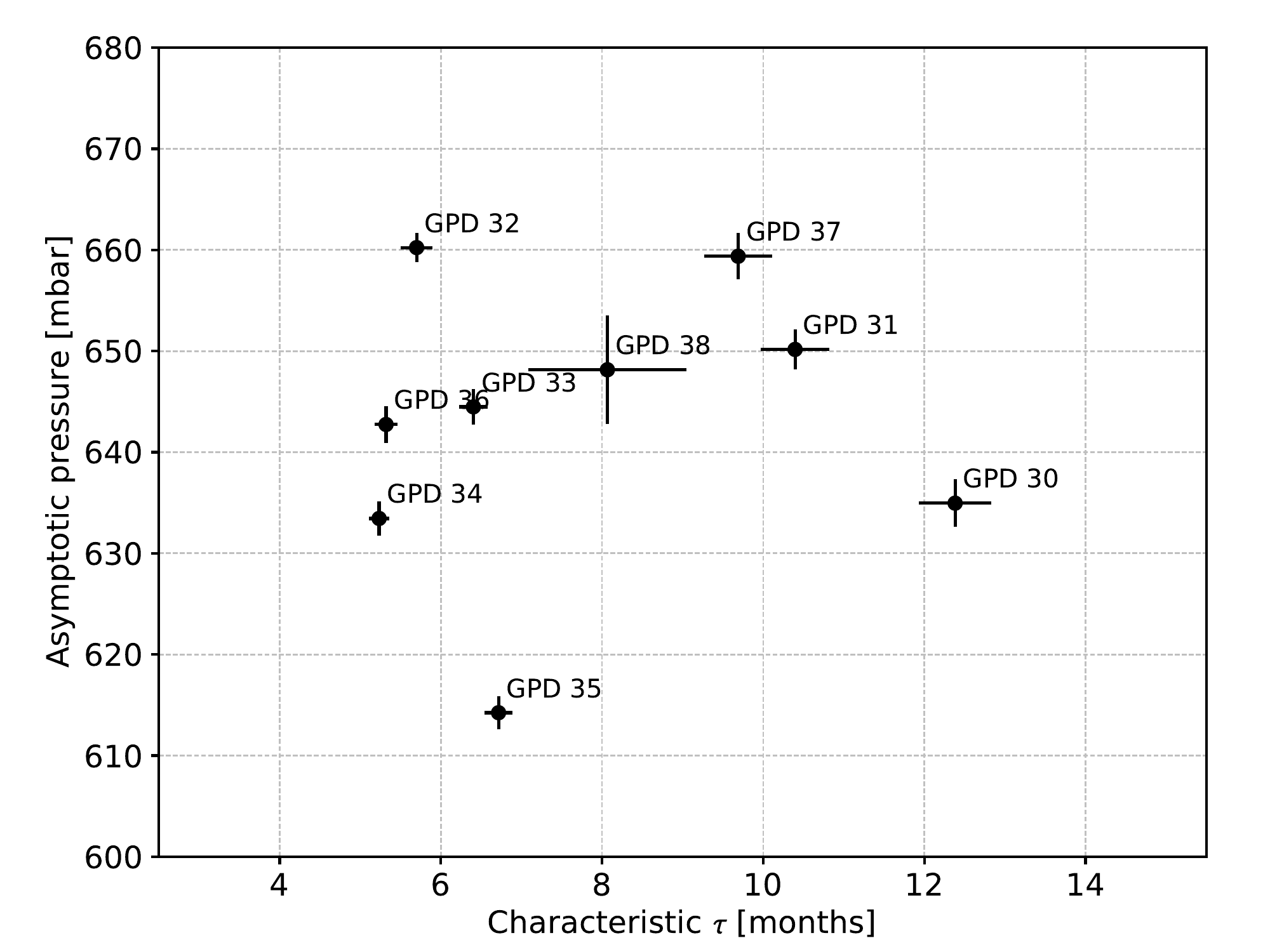}
    \caption{Top-level summary of the best-fit parameters for the pressure models for
    a compilation of detectors.}
    \label{fig:secular_parameter_summary}
\end{figure}

The objective function
that we minimize is
\begin{align}
   \chi^2(\tau, \Delta_p, C_L, C_G, C_Q) =
   \sum_{i=1}^{n_L} \frac{\left(L_i - 
   C_L L\left(p\left(t_i; \tau, \Delta_p\right)\right)\right)^2}{\sigma_{L_i}^2 + \sigma_{L_{sys}}^2} +
   \nonumber\\ 
   \sum_{j=1}^{n_G} \frac{\left(G_i -
   C_G G(p(t_i; \tau, \Delta_p))\right)^2}{\sigma_{G_i}^2 + \sigma_{G_{sys}}^2} +
   \sum_{k=1}^{n_Q} \frac{\left(Q_i -
   C_Q Q(p\left(t_i; \tau, \Delta_p\right)\right)^2}{\sigma_{Q_i}^2 + \sigma_{Q_{sys}}^2}.
\end{align}
Note that we simultaneously fit three scale parameters for the three proxies to allow for 
modeling uncertainties or setup-dependent scaling. These fit byproducts can be used to 
evaluate the consistency of the modeling across different detectors. Also note that,
since most of the measurements are statistics-limited, we allow for a constant systematic
error on each of the three proxies.
Figure~\ref{fig:secular_parameter_summary} shows a summary of the fit parameters.

Finally, we report independent, quantitative evidence that the pressure inside the gas cell decreases using data from the standard detector metrology that is performed to support the
alignmnent of the detector units with the mirror-module assemblies in the integration stage.
Prior to filling, the vertical position of the center of the Be window is measured for
each detector with respect to the bottom plane of the titanium frame. In this
configuration the differential pressure between the two sides of the window is zero,
and therefore the window itself is, in principle, perfectly aligned to the titanium
bottom plane. This first measurement constitutes the zero for the following ones.

After the gas filling, the window is subjected to a $\sim 200$~mbar differential pressure
(that, as we have seen, is changing with time), which causes a measurable movement of the
vertical position of the window center. Based on FEM simulations and actual tests, this
shift is expected to be linear with the differential pressure 
up to several hundreds mbar:
\begin{align}
    \Delta z \propto p_\text{atm} - p_\text{in}.
\end{align}
It is therefore possible, at least in principle, to gauge the internal pressure in the gas
cell from a direct metrological measurement. A precision of $\sim 5~\mu$m can be 
achieved with a CMM with an optical head.

\begin{figure}[htbp!]
    \centering
    \includegraphics[width=\linewidth]{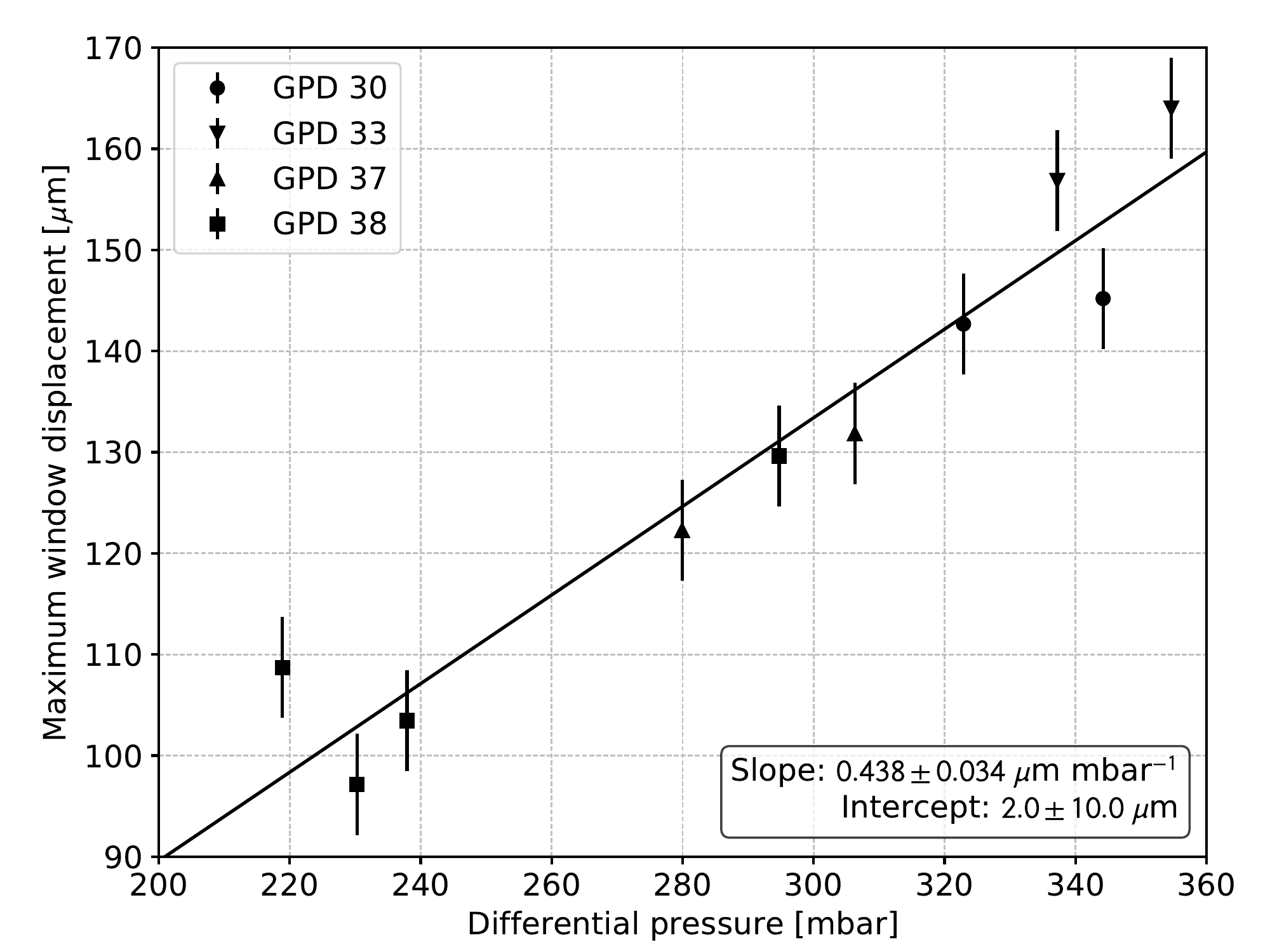}
    \caption{Compilation of window displacement measurements for the GPD in the
    flight control sample, at different moments in time, as a function of the pressure
    in the gas cell, estimated from our combined fit. The excellent linearity provides
    quantitative support for the goodness of the entire fitting process.}
    \label{fig:be_window_displacement}
\end{figure}

In figure~\ref{fig:be_window_displacement} the window displacement measurement is shown
for each GPD in the flight control sample, and at different moments in time,
as a function of the pressure in the gas cell, estimated from the combined fit described above.
Different points in the plot with the same color track directly the pressure evolution of
any given detector. Overall, the data points are consistent, as expected, with a straight
line, for an overall excursion of $50~\mu$m over 
an estimated maximum pressure range in excess of $100$~mbar.
We emphasize that the intercept of the best-fit straight line is consistent with $0$,
as predicted by our naive linear model.

\bibliographystyle{unsrt}
\bibliography{bibliography}

\end{document}